\documentclass[aps,reprint,floatfix,showkeys]{revtex4-2}

\usepackage{graphicx}% Include figure files
\usepackage{dcolumn}% Align table columns on decimal point
\usepackage{bm}% bold math
\usepackage{amsmath}
\usepackage{bbold}
\renewcommand{\vec}[1]{\boldsymbol{#1}}
\usepackage{csquotes}
\usepackage{booktabs}
\usepackage[hidelinks]{hyperref}
\usepackage{nicefrac}

\usepackage{float}
\usepackage{placeins}

\usepackage{multirow}

\usepackage{siunitx}[]
\usepackage{tikz}
\usetikzlibrary{math}
\usetikzlibrary{shapes.geometric}
\usetikzlibrary{arrows}
\usetikzlibrary{positioning}
\usetikzlibrary{shapes}
\usetikzlibrary{intersections}
\usetikzlibrary{arrows.meta}

\usepackage{pgfplots}
\usepgfplotslibrary{groupplots}
\pgfplotsset{compat=newest}
\pgfplotsset{plot coordinates/math parser=false} 
\usetikzlibrary{pgfplots.colorbrewer}
\usepgfplotslibrary{colormaps}
\pgfplotsset{colormap/violet}
\pgfplotsset{colormap/viridis}
\pgfplotsset{
    colormap={myplasma}{
        rgb255=(68, 1, 84);
        rgb255=(59, 15, 112);
        rgb255=(140, 41, 129);
        rgb255=(222, 73, 104);
        rgb255=(252, 166, 54);
        rgb255=(240, 249, 33)
    },
}
\usepgfplotslibrary{external} 
\tikzset{external/up to date check=md5}

\newcommand{\red}[1]{\textcolor{black}{#1}}

\usepackage{glossaries}

\usepackage[]{tikz}
\usepackage{tikzscale}

\newacronym{dem}{DEM}{discrete element method}

\definecolor{brickred}{rgb}{0.8, 0.25, 0.33}
\definecolor{darkorange}{rgb}{1.0, 0.55, 0.0}
\definecolor{persiangreen}{rgb}{0.0, 0.65, 0.58}
\definecolor{persianindigo}{rgb}{0.2, 0.07, 0.48}
\definecolor{cadet}{rgb}{0.33, 0.41, 0.47}
\definecolor{turquoisegreen}{rgb}{0.63, 0.84, 0.71}
\definecolor{sandybrown}{rgb}{0.96, 0.64, 0.38}
\definecolor{blueblue}{rgb}{0.0, 0.2, 0.6}
\definecolor{ballblue}{rgb}{0.13, 0.67, 0.8}
\definecolor{greengreen}{rgb}{0.0, 0.5, 0.0}
\definecolor{razzmatazz}{rgb}{0.89, 0.15, 0.42}
\definecolor{ultramarine}{rgb}{0.07, 0.04, 0.56}
\definecolor{midnightgreen}{rgb}{0.0, 0.29, 0.33}
\definecolor{lavenderpurple}{rgb}{0.59, 0.48, 0.71}
\definecolor{bittersweet}{rgb}{1.0, 0.44, 0.37}
\definecolor{amaranth}{rgb}{0.9, 0.17, 0.31}
\definecolor{forestgreen}{rgb}{0.13, 0.55, 0.13}

\definecolor{myX}{RGB}{102,194,165}
\definecolor{myY}{RGB}{252,141,98}
\definecolor{myZ}{RGB}{141,160,203}

\definecolor{c1}{rgb}{0.7068574918274737, 0.11027871818526241, 0.2747061222663145}%
\definecolor{c2}{rgb}{0.6780437401750237, 0.1857033496010309, 0.10475664313058389}%
\definecolor{c3}{rgb}{0.5819819292481591, 0.28135510723834917, 0.10365807489974799}%
\definecolor{c4}{rgb}{0.5234892407222805, 0.3183040932830017, 0.10308831855626377}%
\definecolor{c5}{rgb}{0.4801916866157751, 0.3399605081271155, 0.10271364194308724}%
\definecolor{c6}{rgb}{0.44359832760663015, 0.3553988979685888, 0.10242748858902176}%
\definecolor{c7}{rgb}{0.40913672397205286, 0.36794412723413256, 0.10218299803254591}%
\definecolor{c8}{rgb}{0.37320211354661986, 0.37925108836394167, 0.10195327596798023}%
\definecolor{c9}{rgb}{0.33135856975649947, 0.3904277497779026, 0.10171733001337946}%
\definecolor{c10}{rgb}{0.2751398809612443, 0.40252912490464393, 0.10145175401497963}%
\definecolor{c11}{rgb}{0.17705545280426335, 0.4169982861326329, 0.10112016988851782}%
\definecolor{c12}{rgb}{0.10370113411519366, 0.41987331701008007, 0.20784080256789766}%
\definecolor{c13}{rgb}{0.10672620229256541, 0.415683426104876, 0.2824387977867041}%
\definecolor{c14}{rgb}{0.10896052083074859, 0.412482459879469, 0.3262929409150867}%
\definecolor{c15}{rgb}{0.11082049183583692, 0.4097465781098524, 0.3585658834499722}%
\definecolor{c16}{rgb}{0.1125289398766243, 0.40717495121262287, 0.38576967807315987}%
\definecolor{c17}{rgb}{0.11424591411818866, 0.4045324469356048, 0.41127417284765605}%
\definecolor{c18}{rgb}{0.11613475317261168, 0.4015563987219263, 0.4376207984793854}%
\definecolor{c19}{rgb}{0.11843118069754927, 0.3978377246048391, 0.46769849466738594}%
\definecolor{c20}{rgb}{0.1215889511379443, 0.3925371130865723, 0.5062751486798138}%
\definecolor{c21}{rgb}{0.12675667510032929, 0.38336665224260497, 0.564172135389151}%
\definecolor{c22}{rgb}{0.13823697094516182, 0.36050804998003744, 0.6775165660716256}%
\definecolor{c23}{rgb}{0.2929229731919379, 0.2517707788576924, 0.9106622939003164}%
\definecolor{c24}{rgb}{0.5044517963791875, 0.15823641315504755, 0.8374462940274711}%
\definecolor{c25}{rgb}{0.5743505600217842, 0.14563681653078617, 0.7277475942695497}%
\definecolor{c26}{rgb}{0.6114384687170589, 0.13753415571759905, 0.6502354669913063}%
\definecolor{c27}{rgb}{0.6363457404298491, 0.13141740987926892, 0.586320806947322}%
\definecolor{c28}{rgb}{0.6557132733878439, 0.12622710616005786, 0.5268534496956088}%
\definecolor{c29}{rgb}{0.6725336221941806, 0.12137095721249477, 0.4649001891758376}%
\definecolor{c30}{rgb}{0.688593421429631, 0.11639483989072555, 0.39157406584043325}%

\definecolor{col1}{rgb}{68, 1, 84}%
\definecolor{col2}{rgb}{65, 68, 135}%
\definecolor{col3}{rgb}{42, 120, 142}%
\definecolor{col4}{rgb}{34, 168, 132}%
\definecolor{col5}{rgb}{122, 209, 81}%
\definecolor{col6}{rgb}{253, 231, 37}%

\usepackage{ifthen}
\newcommand{\Bilder}{ja}
\newcommand{\printfig}[1]{ \ifthenelse{\equal{\Bilder}{ja}}{#1}{\includegraphics[width=0.8\columnwidth]{dummy.png}} }

\begin{document}

\title{Structural features of jammed-granulate metamaterials}

\author{Holger G\"{o}tz}
\affiliation{Institute for Multiscale Simulation, Universit\"{a}t Erlangen-N\"{u}rnberg, Cauerstra\ss{}e 3, 91058 Erlangen, Germany}

\author{Thorsten P\"{o}schel}
\affiliation{Institute for Multiscale Simulation, Universit\"{a}t Erlangen-N\"{u}rnberg, Cauerstra\ss{}e 3, 91058 Erlangen, Germany}

\author{Olfa D'Angelo}
\email{olfa.dangelo@fau.de}
\affiliation{Institute for Multiscale Simulation, Universit\"{a}t Erlangen-N\"{u}rnberg, Cauerstra\ss{}e 3, 91058 Erlangen, Germany}

\date{\today}

\begin{abstract}
  Granular media near jamming exhibit fascinating properties, which can be harnessed to create jammed-granulate metamaterials:
materials whose characteristics arise not only from the shape and material properties of the particles at the microscale, but also from the geometric features of the packing.
For the case of a bending beam made from jammed-granulate metamaterial, we study the impact of the particles' properties on the metamaterial's macroscopic mechanical characteristics. We find that the metamaterial's stiffness emerges from its volume fraction, in turn originating from its creation protocol; its ultimate strength corresponds to yielding of the force network. 
  In contrast to many traditional materials, we find that macroscopic deformation occurs mostly through affine motion within the packing, aided by stress relieve through local plastic events, surprisingly homogeneously spread and persistent throughout bending.
\keywords{jamming; soft robotics; granular metamaterials; micro-to-macro; DEM simulation}
\end{abstract}

\maketitle

\section{Introduction}

When granular materials undergo a jamming transition, their mechanical properties change drastically. 
Through this transition, the granulate changes from a liquid-like, deformable or flowing state, in which it can plastically deform, to a solid-like state, in which relative particles' positions are persistent and the granulate can withstand a finite load before yielding~\cite{Liu2001, OHern2003}. 
From this property of granular media, a range of applications emerges -- most notably their use in soft robotics. 
Jamming-based actuators are remarkable for their range of mechanical properties: from soft and deformable objects, they can be actuated into full rigidity, becoming suitable even for load-bearing applications~\cite{Fitzgerald2020}.

We investigate the macroscale mechanical properties of the granular jammed state. 
More precisely, we consider a granulate in its jammed state as a metamaterial. 
Metamaterials are artificially engineered to have properties not found in naturally occurring materials~\cite{pendryNegativeRefractionMakes2000}. They draw their properties not only from that of their constituents, but also from their structural features.
Thus, the mechanical characteristics of jammed-granulate metamaterials are not only due to the particles' properties on the microscale (such as stiffness, Poisson ratio, surface friction, size distribution, etc.), but also due to structural features of the metamaterial.
The structure of a jammed granulate is in turn defined by the contact network and its statistics; 
besides particles' properties~\cite{Loeve2010, jiangRoboticGranularJamming2014a, gotzGranularMetaMaterialViscoelastic2022},
it follows from the details of the creation of the jammed state, that is, its history~\cite{Chaudhuri2010, Ciamarra2010, Otsuki2012, Hermes2010, Baranau2014, Torquato2010, Kumar2016} and external forces such as gravity~\cite{DAngelo2022}.

The critical volume fraction, $\phi_\text{J}$, at which the jamming transition occurs, and the corresponding coordination number, \red{$z_\text{J}$}, depend on the properties of the particles~\cite{OHern2003, Hermes2010, VanHecke2009} -- in particular, interparticle friction~\cite{VanHecke2009, Ciamarra2011, Mari2014, Bi2011}.
These quantities are among the most significant structural characteristics of the jammed state, 
as the macroscopic properties of  jammed granular media scale with the deviation from \red{$z_\text{J}$}~\cite{OHern2003,majmudarJammingTransitionGranular2007, VanHecke2009}. 
Indeed, the macroscopic properties of the material, such as stiffness, shear strength, yield strength, and others, are mediated by the properties of the contacts between the particles at the microscale.
Therefore, to understand and optimize jammed-granulate metamaterials,
it is insufficient to know their dependence on the microscopic particle properties. 
Such studies can be  found in the literature, e.g., for applications in robotics~\cite{Brown2010, Goetz2022,jiangVariableStiffnessGripper2019, Santarossa2023}, construction \cite{Huijben2011, Huijben2014, Brigido2022}, or medical technologies \cite{cianchettiSTIFFFLOPSurgicalManipulator2013, Loeve2010}. 

In the current paper, we focus on the structural aspect of jammed-granulate metamaterials. 
To identify their material characteristics, we numerically perform four-points bending on a metamaterial beam, composed of a granular medium enclosed by an elastic membrane pressure-jamming the granulate.
The beam's response to bending is largely determined by the granulate's compressive behavior, due to contact forces;
tensile forces are due to the membrane enclosing the granulate~\cite{Huijben2011, Huijben2014, Brigido2022}. 
The share of particles under compression is determined by the positions of the neutral axis, which depends on the applied confining pressure, and on the deformation of the beam (i.e., the strain or stress applied)~\cite{brigido-gonzalezSwitchableStiffnessMorphing2019, Huijben2014}.
We vary the Young's modulus and friction of the particles, and study the 
macroscopic stiffness and ultimate strength of the metamaterial, before focusing on the underlying structural features of the contact network.

\section{System Description}
\label{sec:system}

\subsection{Setup}
\label{sec:system:setup}

\begin{figure}[ht]
  \centering
%  \printfig{\input{media/img/simSetup/BeamSketch}}
\includegraphics{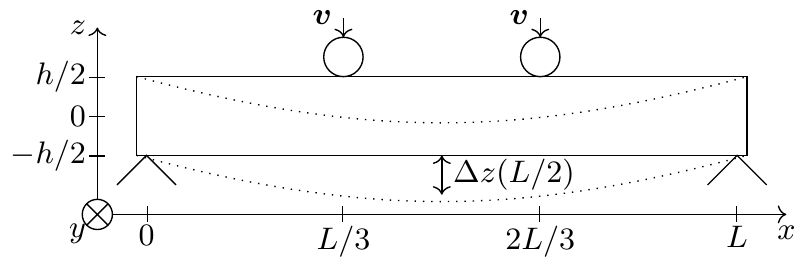}
  \caption{Sketch of the system studied: a simply supported beam subjected to four-point bending.}
  \label{fig:sketch_simply_supported_beam}
\end{figure}

We consider a jammed-granulate metamaterial beam of rectangular cross-section ($4\,\si{\cm} \times 8$\,\si{\cm}) and length $L=\SI{40}{\cm}$ (see Fig.~\ref{fig:sketch_simply_supported_beam}). 
The beam consists of granular particles confined by a soft membrane. In numerical simulations, the beam is subjected to four-point bending. 

Four-point bending produces constant shear stress between the loading points and is, therefore, the preferred test to determine material characteristics. Two rods were initially placed above the beam at positions $L/3$ and $2L/3$, as per industry standards~\cite{ASTM2020_bending, DINENISO2011_bending}. They move downwards at constant velocity $\SI{0.2}{\mm\per\s}$, thus imposing bending. The deflection $\Delta z $ at position $L/2$ quantifies the beam's deformation.
\red{Perturbations propagate at the speed of sound in granular media, which has been estimated to be at least $\SI{50}{\meter\per\second}$~\cite{liuSoundGranularMaterial1993,tellAcousticWavesGranular2020}.
A rough calculation, even for our softest particles, gives the same order of magnitude, $\mathcal{O}(\SI{10}{\meter\per\second})$.
This is much higher than the applied deformation velocity of $\mathcal{O}(10^{-4}\,\si{\meter\per\second})$.
Hence, the applied deformation is quasi-static.}

We perform discrete-elements (DEM) simulations \cite{poschelComputationalGranularDynamics2005, matuttisUnderstandingDiscreteElement2014, ludingIntroductionDiscreteElement2008} using MercuryDPM \cite{thorntonRecentAdvancesMercuryDPM2023}. The membrane is modelled by a mass-spring system with triangular elements for granulate-membrane interactions as introduced recently~\cite{goetzDEMSimulationThinElastic2022}.  The details of the simulation method can be found in Appendix~\ref{sec:methods}.

\subsection{Initial conditions: Preparation protocol}

Granular packings are largely determined by their formation procedure.
We use the coefficient of friction between particles as a parameter to generate jammed-granulate packings of different initial packing characteristics. 

\begin{figure*}[htb]
     \centering
%  \printfig{ \input{media/img/simSetup/schematic_procedure.tex}}  
\includegraphics{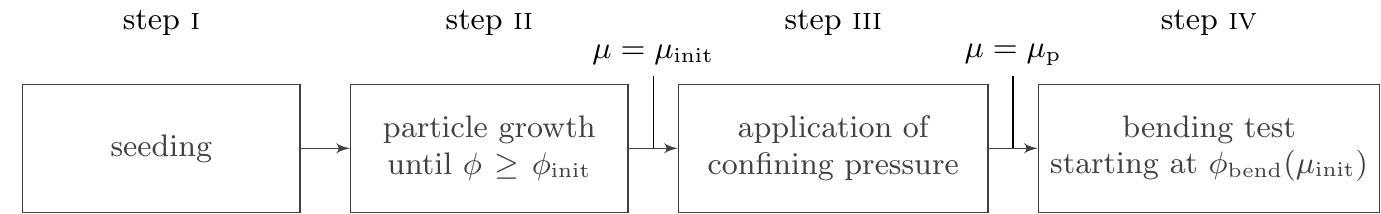}
  \caption{Simulation protocol. For discussion see body of text.}
 \label{fig:schematic_procedure}
\end{figure*}

The  preparation protocol comprises the following steps:
\begin{enumerate}
\item \label{en:step1}
\textit{Initial placement:}
frictionless particles ($\mu=0$) with uniformly distributed radii $R_i\in[1.66, 1.84]\,\text{mm}$ are placed at random positions inside the membrane, such that they do not enter in contact with one another.
\item \label{en:step2}
\textit{Lubachevsky-Stillinger:}
to generate a packing with a defined volume fraction $\phi_\text{init}$, we apply the Lubachevsky-Stillinger algorithm \cite{lubachevskyGeometricPropertiesRandom1990}: 
the particles' radii increase gradually while keeping the membrane's interior volume constant until a desired volume fraction $\phi_\text{init}$ is reached. Note that the volume fraction $\phi_\text{init}$ is that \emph{before} applying pressure to jam the packing. The entire system is contained in a cuboid-shaped solid mold to ensure that the beam keeps its shape and does not increase in size by deforming the elastic membrane.

\item \label{en:step3}
\textit{Application of confining pressure:}
we gradually apply a pressure difference $\Delta p$ between the beam's interior and the ambient surrounding to the confining membrane (effectively applying vacuum inside the beam). $\Delta p$ is termed \emph{confining pressure} in the following.
For this step, we apply the friction coefficient $\mu=\mu_\text{init}$. The jammed-granulate beam shrinks in volume, correspondingly.
While $\Delta p$ is gradually increased, the cuboidal shape of the granular beam is preserved by moving the planes constituting the solid mold, such that the membrane remains in contact with the mold.
\end{enumerate}

The choice of the friction coefficient $\mu_\text{init}$ determines the volume fraction, $\phi_\text{bend}$, at the beginning of the bending test.
After initialization and before bending, we set the particles' friction coefficient to $\mu=\mu_\text{p}$,
the particles' friction during the bending test. 
Figure~\ref{fig:schematic_procedure} summarizes the full simulation protocol.

\subsection{Initial conditions: Characterization}
\subsubsection{Average coordination number}

\begin{figure}[h!]
  \centering  
%  \printfig {\input{media/img/initStudy/packingProperties.tex} }
\includegraphics{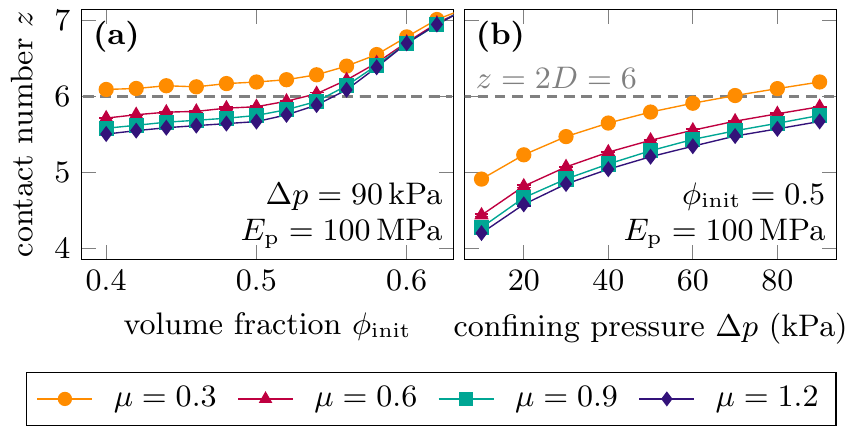}
  \vspace{-15pt}
\caption{Average contact number, $z$, of the packing generated according to the preparation protocol. (a) $z$ as a function of the initial packing fraction, $\phi_\text{init}$, for fixed confining pressure $\Delta p=\SI{90}{\kilo\pascal}$ and several values of the initial friction coefficient, $\mu_\text{init}$. (b) $z(\Delta p)$ with fixed parameters $\phi_\text{init} = 0.5$ and $\mu_\text{init}$ as per legend.}
  \label{fig:packing:packingPressureContact}
\end{figure}

An important characteristic of a jammed packing is the average number of contacts \red{per particle}, $z$. Since the preparation protocol determines the properties of the packing, the average contact number should depend on the parameters of this protocol, namely: the volume fraction $\phi_\text{init}$ before application of confining pressure; the friction coefficient $\mu_\text{init}$ during the application of the pressure; the magnitude of the confining pressure $\Delta p$. 
Figure~\ref{fig:packing:packingPressureContact}  shows these functions: $z(\phi_\text{init})$ with fixed parameters $\Delta p$ and $\mu_\text{init}$, and $z(\Delta p)$ with fixed parameters $\phi_\text{init}$ and $\mu_\text{init}$.

In both figures, we see a decrease in average contact number with increasing friction coefficients.
Similar behavior is known from the theory of hard spheres. 
Here, asymptotically, the average number of contacts scales as $z=2\,D$ for frictionless particles and as $z=D+1$ for $\mu\to\infty$, where  $D$ is the dimension of the system. For finite friction $\mu$, $z$  ranges between these limits, which can be described by a constraint function (see \cite{Song2008} for details). 

\subsubsection{Pair correlation}
\label{sec:pairCorrelation}
\begin{figure}[hbt!]
  \centering
%  \printfig{\includegraphics[width=\linewidth]{media/img/initStudy/pairCorrelation.tikz}}
%  \printfig{\input{media/img/initStudy/pairCorrelation.tex}}
\includegraphics{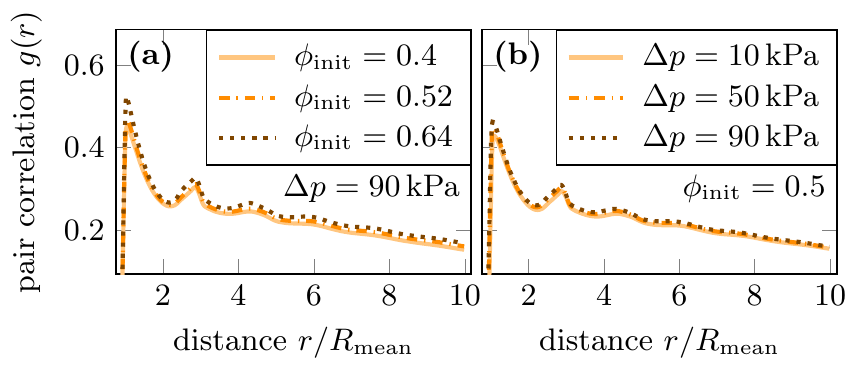}
  \caption{Radial pair correlation function for (a)~fixed confining pressure, $\Delta p=\SI{90}{\kilo\pascal}$, and various values of the initial packing fraction, $\phi_\text{init}$, and (b) fixed initial packing fraction, $\phi_\text{init}=0.5$, for various values of confining pressure. Particle material parameters: $E_\text{p} = \SI{100}{\mega\pascal}$ and $\mu_\text{p} = 0.3$.}
  \label{fig:packing:pairCorrelation}
\end{figure}

The pair correlation function
\begin{equation}
  g(r) = \left\langle \frac{n_i(r)}{4\pi (r/r_\text{mean})^2} \right\rangle_i
\end{equation}
characterizes the regularity of homogeneous particulate material. 
Here, $n_i(r)$ is the number of particles intersecting with the surface of a sphere of radius $r$ placed at the center of particle $i$. $\langle\cdot\rangle_i$ indicates the average over all particles $i$ in the system.
Figure \ref{fig:packing:pairCorrelation} shows $g(r)$ for fixed pressure, $\Delta p=90\,\text{kPa}$, and variable initial packing fraction, $\phi_\text{init}$ (Fig.~\ref{fig:packing:pairCorrelation}a), and for fixed initial packing fraction, $\phi_\text{init}=0.5$, and variable pressure. 
In all cases studied, 
we do not see any sign of crystallization, 
\red{which would appear in the pair correlation function as narrow vertical peaks due to long-range ordering.
Rather, we observe vanishing peaks as $r$ increases, showing only short-range ordering.
This indicates that our system remains amorphous in all cases shown~\cite{jinFirstorderPhaseTransition2010}.
% hence jammed and not crystallized. 
}

\subsection{Beam stiffness and ultimate strength}
The macroscopic response of the beam to bending is its deflection $\Delta z$ (see Fig.~\ref{fig:sketch_simply_supported_beam}) and its resistance to the rods' motion, measured as the force $F$ acting on the rods. The Euler-Bernoulli beam theory \cite{truesdellRationalMechanicsFlexible1960} provides the following relations to convert the measurements into stress $\sigma$ and strain $\epsilon$:
\begin{align}
  \sigma& = \frac{FL}{h^2d}\label{eq:timoshenkoStress},\\
  \epsilon& = \Delta z\frac{108h}{23L^2}\label{eq:timoshenkoStrain}.
\end{align}
Here, the beam's length $L$ (which is also its support span), height $h$, and depth $d$ define the geometry of the beam (see Sec.~\ref{sec:system:setup}). Details on the derivation of this relation are given in Appendix~\ref{apx:timoshenko}.

\begin{figure}[htb!]
%\printfig{  \input{media/img/stressStrainExample/stressStrainExample.tex}}
\includegraphics{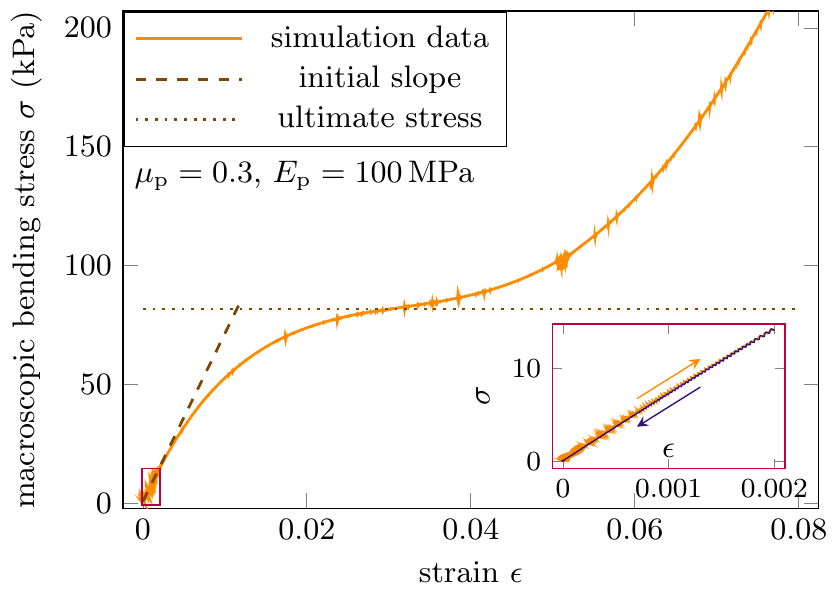}
  \vspace{-7pt}
  \caption{Exemplary stress-strain curve. The elastic modulus (initial slope of the curve, dashed line) and ultimate strength (dotted line) are displayed. The inset shows an examplary stress-strain curve for a cyclic bending simulation up to $\epsilon=0.002$. The insets region is indicated by a red rectangle in the main figure. The displayed data was obtained for a system with $E_\text{p} = \SI{100}{\mega\pascal}$ and $\mu_\text{p} = 0.3$.}
  \label{fig:stressStrainExample}
\end{figure}

The typical stress-strain curve is presented in Fig.~\ref{fig:stressStrainExample}.
It shows three stages: an initial steep increase followed by a plateau or turning point, and, finally, a second steep increase. For the initial steep increase, in the limit of small strain, we assume the linear relation $\sigma=E_\text{b}\epsilon$. Here, the proportionality constant $E_\text{b}$ measures the beam's macroscopic stiffness,
determined by linearly fitting the stress-strain curve for $\epsilon\leq0.002$. 
Note that the term \enquote{stiffness} is used, \red{although the beam does not follow the exact load path when released during this phase.}
This property is shared by many ordinary materials, which also experience some structural plasticity even for small deformation, responsible for non-fully recovered properties and aging. \red{Yet, the difference between loading and unloading is low in the considered range, as can be seen in the inset of Fig.~\ref{fig:stressStrainExample}.}

The plateau or turning point indicates the ultimate strength $Y_\text{b}$ of the beam, i.e., the maximum amount of stress, $\sigma$, that the jammed system can resist. 
\red{To calculate $Y_\text{b}$, the measured stress and strain signals are smoothed by applying a Gaussian filter and used to calculate the first derivative $\sigma^\prime(\epsilon)$ with central differences. The derivative is filtered again and used to calculate the second derivative $\sigma^{\prime\prime}(\epsilon)$. The ultimate strength is the stress value at which the second derivative switches sign (from negative to positive).}

After this plateau, in the final region of steep increase, \red{the membrane takes over and dictates the behavior of the stress-strain curve.}
Hence, it no longer provides insight into the behavior of the jammed granular media inside the beam. The simulations run until $\epsilon\approx0.08$ is achieved to ensure that the stress-strain relation has reached the third region.

\red{In the following, the analysis focuses on the particles, fixing the membrane's stiffness and varying the particles' properties. However, the membrane's properties also influence the beam's response: we showed in previous work that stiffer membranes lead to stiffer beams~\cite{gotzGranularMetaMaterialViscoelastic2022}.}

\section{Results and Discussion}
The results presented below are obtained for a relatively low initial volume fraction of $\phi_\text{init}=0.5$, to avoid artificially jamming the packing before applying the confining pressure, $\Delta p$. The confining pressure is fixed at $\Delta p = \SI{90}{\kilo\pascal}$, a relatively high value, because previous studies showed that jammed beam are more stable for large confining pressure \cite{goetzDEMSimulationThinElastic2022,brigido-gonzalezSwitchableStiffnessMorphing2019}.

\subsection{Microscopic to macroscopic properties
  \label{sec:microMacroProtA}}
The particles' properties are varied to determine the relation between the microscopic particle properties and the macroscopic properties of the jammed granular beam. Specifically, we vary the particles' stiffness $E_\text{p} \in [5,\,100]$\,\si{\mega\pascal} and friction coefficients $\mu_\text{p} \in [0,\,1.2]$.

\begin{figure*}[htb!]
    \centering
%    \printfig{  \input{media/img/elasticModulus/heatmapBefore.tex}}
\includegraphics{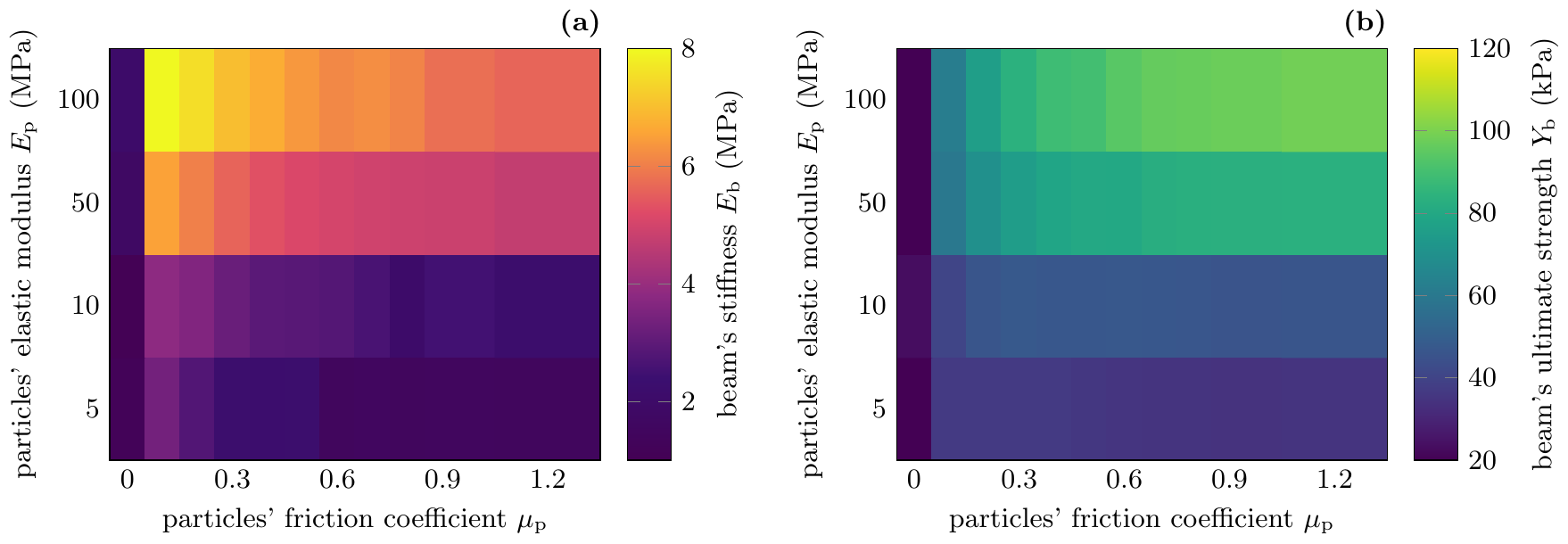}
    \vspace{-12pt}
    \caption{The beam's macroscopic properties as a function of the constitutive particles' microscopic properties. The beam's (a) stiffness modulus, $E_\text{b}$, and (b) ultimate strength, $Y_\text{b}$, are shown for the particles' friction coefficients, $\mu_\text{p} \in [0,\,1.2] $, and the particles' stiffnesses, $E_\text{p} \in [5,\,100]\,\si{\mega\pascal}$, obtained with protocol \textsc{a}: the volume fraction, $\phi_\text{bend}$, depends on the particles' friction coefficient during bending. }
    \label{fig:heatmapElasticModulusProtA}
\end{figure*}

The beam's macroscopic properties are initially explored for the intuitive choice $\mu_\text{init}=\mu_\text{p}$, where the granular media's volume fraction at the beginning of the bending test, $\phi_\text{bend}$, (i.e. the volume fraction \emph{after} the application of the confining pressure) is a function of the particles' friction coefficient during bending: $\phi_\text{bend} = \phi_\text{bend}(\mu_\text{init}) = \phi_\text{bend}(\mu_\text{p})$. The choice $\mu_\text{init}=\mu_\text{p}$ is termed {protocol \textsc{a}} in the following.
Figure \ref{fig:heatmapElasticModulusProtA} displays a heatmap of the beam's macroscopic stiffness $E_\text{b}$ (Fig. \ref{fig:heatmapElasticModulusProtA}a) and ultimate strength $Y_\text{b}$ (Fig. \ref{fig:heatmapElasticModulusProtA}b) for the parameter space of interest.

Stiffer particles lead to a higher beam's macroscopic stiffness 
\red{(Figs.~\ref{fig:heatmapElasticModulusProtA}a, \ref{fig:linePlotStiffness}a)}, \red{which can be expected}.
As shown in previous work \cite{gotzGranularMetaMaterialViscoelastic2022}, the beam's stiffness is related to the particles' elastic modulus following a power law.

One might naively expect that a large friction coefficient also leads to a beam of high stiffness, as highly frictional particle-particle contacts can withstand high load. Surprisingly however, a large friction coefficient \emph{decreases} the stiffness of the beam. Our hypothesis is that this is a result of the lower number of contacts within the jammed packing for large $\mu_\text{p}$: if the confining pressure is applied to a packing of high friction, the particles pack with low density. Therefore, the total number of contacts within the jammed packing is low. This results in each individual contact being under high load, and in turn, in a low macroscopic stiffness of the jammed phase. We will explore this idea further in the next section.

Note that $\mu_\text{p}=0$ represents a special case where no force opposes
particles to sliding against each other, which lets particles reorganize easily. 
The low macroscopic stiffness for $\mu_\text{p}=0$ indicates that 
to develop a significant mechanical resistance to external load, 
friction is necessary for the stability of the jammed packing.
% to be stable against load and show a significant response.

The beam's macroscopic ultimate strength $Y_\text{b}$
\red{(Figs.~\ref{fig:heatmapElasticModulusProtA}b, \ref{fig:linePlotStiffness}b)} 
shows two effects: 
for hard particles ($E_\text{p}=50\,\text{MPa}$ and $E_\text{p}=100\,\text{MPa}$), \red{increasing} $\mu_\text{p}$ leads to \red{an increase} in $Y_\text{b}$, because large forces are necessary to allow particles to slide against each other to relieve load. For soft particles ($E_\text{p}=5\,\text{MPa}$ and $E_\text{p}=10\,\text{MPa}$), 
\red{this is only true at low particles' friction, $\mu_\text{p}<0.3$.}
%large $\mu_\text{p}$ lead to high $Y_\text{b}$, but only if $\mu_\text{p}$ remains low. 
For large $\mu_\text{p}$,
%($\mu_\text{p}>0.3$), 
a further increase of $\mu_\text{p}$ leaves $Y_\text{b}$ unchanged or decreases its value. 
This shows the competition of two effects: a high friction coefficient $\mu_\text{p}$ increases the amount of force \emph{one} contact can carry, but also reduces the number of contacts within the system, because the system jams at lower volume fraction. At large values of $\mu_\text{p}$, further increase of load carried per contact cannot compensate for the reduced number of contacts.

Note that the spread of the ultimate strength is generally smaller for low particle stiffness than for large particle stiffness. This is because the higher deformation and higher density of softer particles hampers the relative motion between neighboring particles. 

\subsection{Influence of preparation through density}
\begin{figure*}[!hbt]
    \centering
%    \printfig{\input{media/img/elasticModulus/heatmapAfter.tex}}
\includegraphics{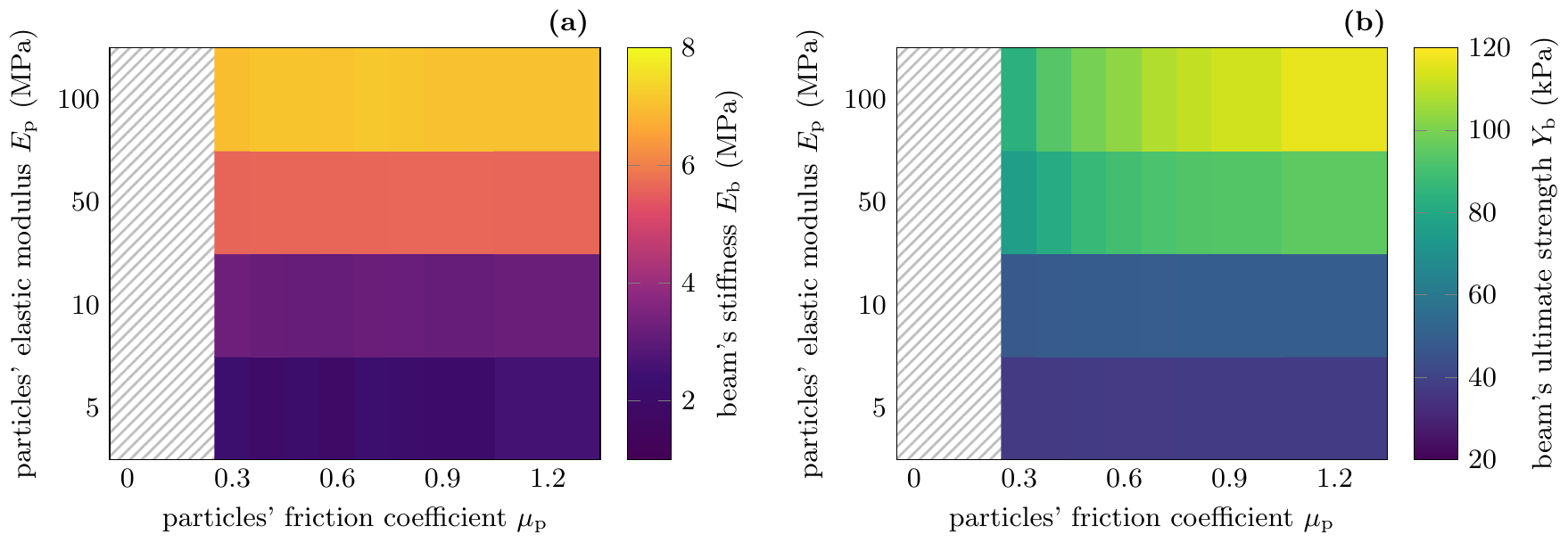}
    \vspace{-12pt}
    \caption{The beam's macroscopic properties as a function of the constitutive particles' microscopic properties. The beam's (a) stiffness modulus, $E_\text{b}$, and (b) ultimate strength, $Y_\text{b}$, are shown for the particles' friction coefficients, $\mu_\text{p} \in [0.3,\,1.2] $, and the particles' stiffnesses, $E_\text{p} \in [5,\,100] $, obtained with protocol \textsc{b}: the volume fraction, $\phi_\text{bend}$, is the same for all values of the particles' friction coefficient, $\mu_\text{p}$. }
    \label{fig:heatmapElasticModulusProtB}
\end{figure*}  

To eliminate the effect of the packing's density and its number of contacts, the bending process is performed with a fixed value of $\mu_\text{init}=0.3$ but several $\mu_\text{p}$. This preparation protocol is termed {protocol \textsc{b}} in the following. Here, the granular medium's volume fraction $\phi_\text{bend}\equiv\phi_\text{bend}(\mu_\text{init})$ at the onset of bending is constant and independent of the particles' friction during bending, $\mu_\text{p}$. 
Values of $\mu_\text{p}<\mu_\text{init}$ are excluded for protocol \textsc{b}, because lower friction coefficients allow for further compression of the beam and thus $\phi_\text{bend}$ would no longer be independent of $\mu_\text{p}$. Figure~\ref{fig:heatmapElasticModulusProtB} displays the heatmap of the beam's macroscopic stiffness $E_\text{b}$ (Fig.~\ref{fig:heatmapElasticModulusProtB}a) and ultimate strength $Y_\text{b}$ (Fig.~\ref{fig:heatmapElasticModulusProtB}b) obtained with protocol \textsc{b}.

The beams' macroscopic stiffness $E_\text{b}$ is equal for equal values of $E_\text{p}$. The particles' friction coefficient $\mu_\text{p}$ does not influence $E_\text{b}$. As hypothesized above, the effect seen for protocol \textsc{a}, where high friction leads to low macroscopic stiffness, is a result of different contact numbers achieved for different friction coefficients: more contacts provide greater resistance for the same amount of deformation. 
The macroscopic stiffness of the jammed phase is defined by its volume fraction through the number of contacts, as each contact can bear a finite amount of the load. But the inter-particle friction does not otherwise influence the macroscopic stiffness of the jammed packing.

Contrariwise, the beam's macroscopic ultimate strength $Y_\text{b}$ increases with increasing $\mu_\text{p}$ (Fig. \ref{fig:heatmapElasticModulusProtB}b). This contrasts with protocol \textsc{a}, where an increase in already large $\mu_\text{p}$ decreases $Y_\text{b}$ for systems with low particle stiffness.
The high ultimate strength for large $\mu_\text{p}$ results from high frictional forces, which need to be overcome for the beam to yield. Systems obtained with protocol \textsc{b} have equal density $\phi_\text{bend}$, hence, the system is not weakened by a low amount of contacts for large $\mu_\text{p}$.
\begin{figure*}[!hbt]
    \centering
%  \printfig{  \input{media/img/elasticModulus/eyLinePlotHorizontal}}
\includegraphics{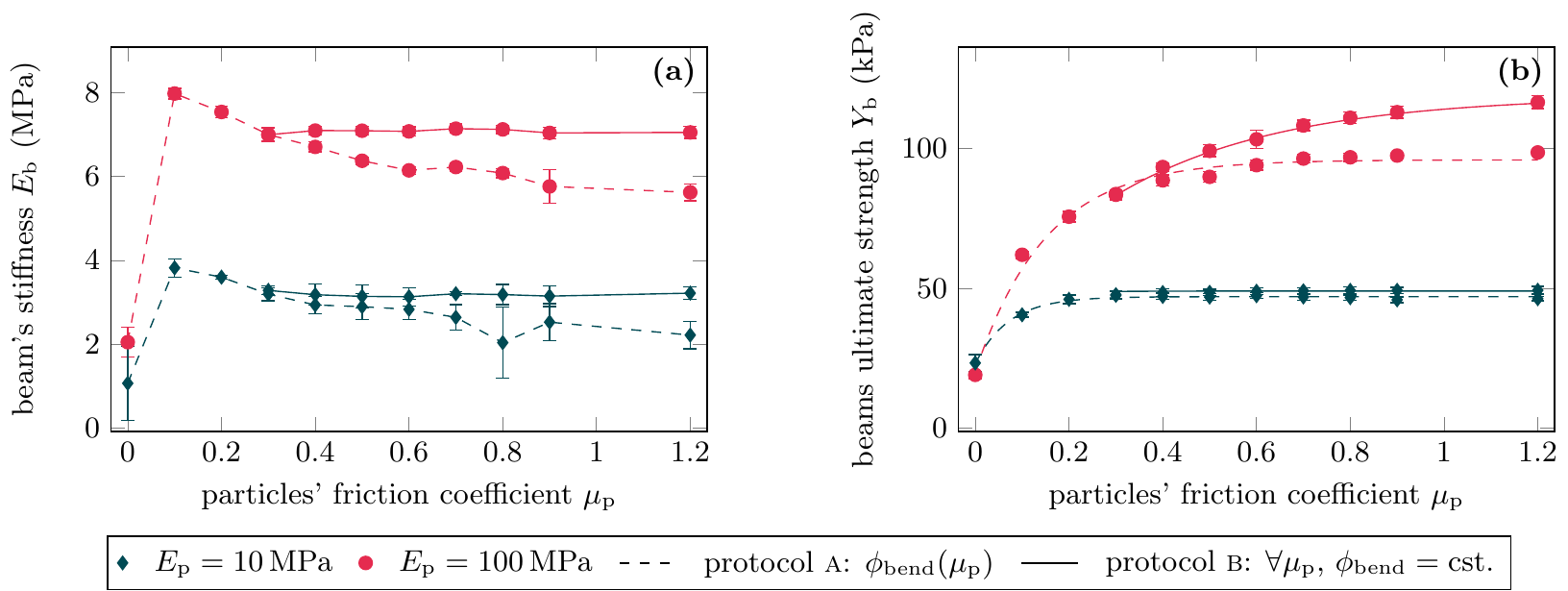}
    \caption{Macroscopic mechanical properties of the beam for increasing particles' friction coefficients, $\mu_\text{p} \in [0, \, 1.2]$. (a) Macroscopic beam's stiffness modulus, $E_\text{b}$, lines connect the points. (b) Macroscopic beam's ultimate strength, $Y_\text{b}$, lines indicate an exponential fit.}
    \label{fig:linePlotStiffness}
    \label{fig:linePlotUltimateStrength}
  \end{figure*}

Figure \ref{fig:linePlotStiffness} shows a direct comparison of the protocols \textsc{a} and \textsc{b}. For $\mu_\text{p}=0.3$, where $\phi_\text{bend}$ is similar in both protocols, identical macroscopic beam properties are observed. For $\mu_\text{p}>0.3$, both the beam's macroscopic stiffness $E_\text{b}$ (Fig.~\ref{fig:linePlotStiffness}a) and ultimate strength (Fig.~\ref{fig:linePlotStiffness}b) are higher for protocol \textsc{b} than for protocol \textsc{a}, because protocol \textsc{b} leads to higher density $\phi_\text{bend}$. 
That is, if $\phi_\text{bend}$ is forced to be constant (protocol \textsc{b}), $\mu_\text{p}$ has no influence on the beam's stiffness, $E_\text{b}$, while $Y_\text{b}$ increases with increasing $\mu_\text{p}$. \red{In both protocols, the tunability of $E_\text{b}$ and $Y_\text{b}$ with particles' friction $\mu_\text{p}$ is weak. If present at all, it is strongest for small friction, $\mu_\text{p} \to 0$.}
An increase of particle stiffness, $E_\text{p}$, directly correlates with an increase of the beam's macroscopic stiffness, $E_\text{b}$, and ultimate strength, $Y_\text{b}$.

Figure~\ref{fig:linePlotStiffness}b suggests that the beam has a finite ultimate strength, $Y_\text{b}$, even in the limit of $\mu_\text{p}\to\infty$. To approximate this behavior, we fit an exponential function of the form:
\begin{equation}
  Y_\text{b}(\mu_\text{p}) = \left(Y_\text{b}^0-Y_\text{b}^\infty \right)e^{-a\mu_\text{p}} + Y_\text{b}^\infty.\label{eq:fitYb}
\end{equation}
% Here, $Y_\text{b}(0)\equiv Y_\text{b}(\mu_\text{p}=0)$ is the beam's ultimate strength for zero particles' friction and $Y_\text{b}(\infty)\equiv Y_\text{b}(\mu_\text{p}\to\infty)$ is the beam's ultimate strength in the limit of an infinitely large particles' friction coefficient. 
Here, $Y_\text{b}^0\equiv Y_\text{b}(\mu_\text{p}=0)$ and $Y_\text{b}^\infty\equiv Y_\text{b}(\mu_\text{p}\to\infty)$ are the beam's ultimate strength in the limit of zero friction and infinitely large friction coefficients. 
The fit constant $a$ describes the curvature of the fit function. 
Lines in Fig.~\ref{fig:linePlotStiffness}b present the fitted functions. 
Fit-parameters for all simulated $E_\text{p}$ are shown in Table~\ref{tab:expFitParams}.

\begin{table}[ht]
  \centering
  \caption{\label{tab:expFitParams}
  Fit parameters of Equation \eqref{eq:fitYb} ($Y_\text{b}(\mu_\text{p}) $), for different values of particles' elastic modulus, $E_\text{p}$, for both protocols \textsc{a} and \textsc{b}.}
  \begin{tabular}{l l l l l}
  \toprule
   $E_\text{p}$ (\si{\mega\pascal})~~~~ &  protocol~~~~ & $Y_\text{b}^0$ (\si{\kilo\pascal})~~~~ & $Y_\text{b}^\infty$ (\si{\kilo\pascal})~~~~ & a~~~~ \\
   \midrule
    \multirow{2}{*}{$5$}
     & \textsc{a} &  18.31 & 35.78 & 251.52 \\
     & \textsc{b} &  36.01 & 37.64 & 2.53 \\
    \hline
    \multirow{2}{*}{$10$}
     & \textsc{a} &  23.17 & 46.92 & 13.92 \\
     & \textsc{b} &  40.63 & 49.31 & 5.31 \\
    \hline
    \multirow{2}{*}{$50$}
     & \textsc{a} &  18.89 & 81.58 & 9.27 \\
     & \textsc{b} &  29.14 & 95.63 & 3.98 \\
    \hline
    \multirow{2}{*}{$100$}
     & \textsc{a} &  20.63 & 96.15 & 6.67 \\
     & \textsc{b} &  35.65 & 119.41 & 2.85 \\
    \bottomrule
  \end{tabular}
  \end{table}

\subsection{Robustness of the force network}
\label{sec:forceNetworkRobustness.}

\begin{figure}[!hbt]
    \centering
%    \printfig{      \input{media/img/ForcePerContact/ForcePerContact_E100.tex}}
\includegraphics{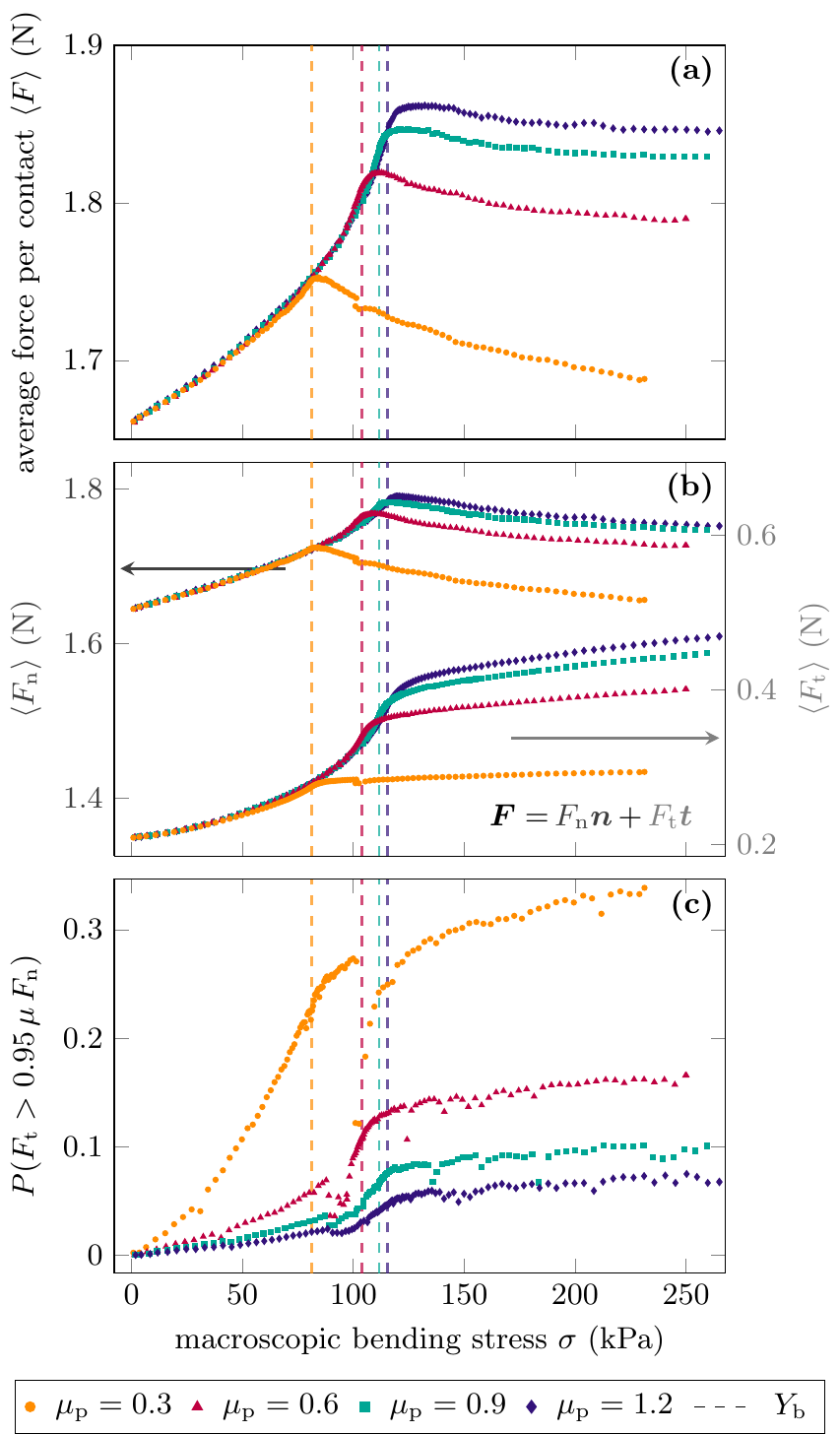}
    \caption{Force per interparticle contact, $F$, as a function of the macroscopic bending stress, $\sigma$, applied to the beam (all measurements follow protocol \textsc{b}, i.e. start at similar volume fraction $\phi_\text{bend}$). The average force per contact $\langle F\rangle$ (a) is broken down into two components: the average normal force $\langle F_\text{n}\rangle$ (b, left axis) and the average tangential force $\langle F_\text{t}\rangle$ (b, right axis). The fraction of contacts close to or at the Coulomb limit indicates the overall fragility of the contact network (c). All variables are shown for hard particles ($E_\text{p}=\SI{100}{\mega\pascal}$), and particles' friction coefficients $\mu_\text{p} \in [0.3,\,0.6,\,0.9,\,1.2]$. Vertical lines drawn at $\sigma=Y_\text{b}(\mu_\text{p})$ indicate the ultimate stress obtained from the stress strain measurement.}
    \label{fig:averageForcePerContact}
\end{figure}

The beam's ability to resist external bending stress originates from the force network formed among the particles. Hence, investigating the force network throughout the bending process can help understanding the mesoscopic meaning of the ultimate strength.

Figure~\ref{fig:averageForcePerContact}a displays the average force per contact $\langle F \rangle$ as a function of the external bending stress $\sigma$. 
The data is displayed for protocol \textsc{b} (all bending experiments start at the same volume fraction), to isolate the effect of different friction coefficients independently of volume fraction effects. Two regimes can be distinguished in the average contact force: for an external bending stress lower than the beams' ultimate strength, $\sigma<Y_\text{b}$, the average force increases with increasing $\sigma$. For higher $\sigma$, the average force decreases with increasing $\sigma$. The ultimate strength corresponds to the point where the force network yields under the applied stress.

While the average force decreases for $\sigma>Y_\text{b}$, the average tangential force, $\langle F_\text{t} \rangle$, displayed in Fig.~\ref{fig:averageForcePerContact}b (right ordinate axis), keeps increasing throughout the whole bending process, though the rate of increase is lowered after reaching the ultimate strength. This suggests that more contacts reach the Coulomb limit where particles can slide against each other. The particles' motion then leads to stress release through breaking contacts and lowered normal forces.

The relative amount of contacts within $5\%$ of reaching the Coulomb limit is shown in Fig.~\ref{fig:averageForcePerContact}c. It increases throughout the process, although at a lower rate for $\sigma>Y_\text{b}$. Hence, for a fixed number of contacts (i.e.~at a given volume fraction), the microscopic particle friction $\mu_\text{p}$, which determines the Coulomb limit, also determines the quantitative threshold when the force network will yield.

Note that this behavior is similar for both hard and soft particles (not shown here), with the difference that the value of the average force per contact is higher for hard particles than for soft particles.
% The latter deform rather than stand high contact forces.
%\todo{Agree with last sentence?}

\subsection{Neutral \red{plane} location}

\begin{figure*}[ht]
  \centering
\includegraphics[]{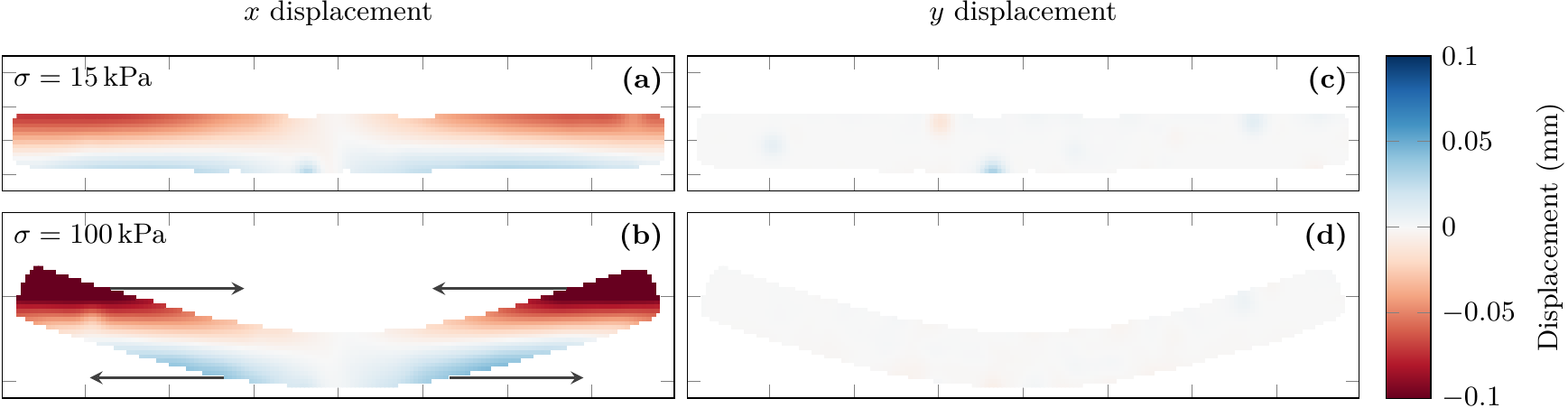}
  \caption{The coarse grained displacement of particles relative to the beam's center. Negative values indicate motion towards the beam's center. Positive values indicate motion away from the beam's center. Subfigures display the displacement in $x$-direction (a, b) and $y$-direction (c, d) at low strain (a, c) and large strain (b, d). The displayed data is obtained from systems with $\mu_\text{p}=0.3$, $E_\text{b}=\SI{100}{\mega\pascal}$, and protocol \textsc{a}. The displacements are measured between two snapshots, 2.5 seconds apart. Details of the calculation are given in Appendix~\ref{supp:displacement}.}
  \label{fig:coarseGrainedDisplacement}
\end{figure*}

Bending divides the beam into two parts: one above the neutral axis, where the beam gets compressed, and one below the neutral axis, where the beam is in tension. This is made obvious by the motion of the particles relative to the beam's center, as shown in Figure~\ref{fig:coarseGrainedDisplacement}. The beam's center is at $(x,\,y)=(L/2,\,0)$ (see also Figure~\ref{fig:sketch_simply_supported_beam}).
In the upper part of the beam, particles are compressed by moving towards the beam's center, while in the lower part of the beam they move away from the beam's center, as the lower part of the beam is in tension. This is in agreement with previous work on jammed-granulate beams~\cite{Huijben2014,brigido-gonzalezSwitchableStiffnessMorphing2019}.

In the upper part of the beam (which is in compression), along the $x$-direction (see Fig.~\ref{fig:sketch_simply_supported_beam}), large displacements are recorded, notably at the beam's extremities. 
At the beginning of the experiment, 
at $\sigma \ll Y_\text{b}$ ($\sigma\lesssim\SI{15}{\kilo\pascal}$, Fig.~\ref{fig:coarseGrainedDisplacement}a, c), all particles move towards the beam's center, but particles at the extremities have to be displaced more to reach the beam's center, which creates the compressive top part. In the later stage of the experiment starting around $\sigma\approx Y_\text{b}$ ($\sigma\gtrsim\SI{100}{\kilo\pascal}$, Fig.~\ref{fig:coarseGrainedDisplacement}b, d), the large displacements at the beam's extremities result from the macroscopic bending motion of the beam as a whole: the extremities are bent inwards by the membrane, which forces this motion.

\red{In $y$-direction, the motion relative to the beam's center is negligible compared to the motion in $x$-direction, as visible in Figs.~\ref{fig:coarseGrainedDisplacement}c, \ref{fig:coarseGrainedDisplacement}d by the absence of color.}

Through the bending test, the neutral plane (zero values in Fig.~\ref{fig:coarseGrainedDisplacement}) shifts towards the top part of the beam.

\subsection{Clusters of motion}

When forced to deform, a confined jammed granular medium responds in two ways: (1) by localized plastic events, where one particle triggers small-scale reorganizations in its surroundings; (2) by large-scale, coordinated affine motion of its individual components, where the beam deforms like a homogeneous material, where all particles move without changing their position relative to their neighbors. 

To investigate the predominant mechanism, we look at the mean squared deviation of particle $i$'s actual displacement relative to its neighboring particles from the displacement it would have, if the packing deformed in an affine manner. It is given by \cite{falkDynamicsViscoplasticDeformation1998}
\begin{align}
  \begin{split}
  D^2(i,\, t,\, t^\prime) = \frac{1}{N}\sum_j \biggl[&\vec{r}_j(t^\prime)-\vec{r}_i(t^\prime) \\
                                                    &- (\mathbb{1} + \vec{\epsilon})(\vec{r}_j(t) - \vec{r}_i(t))\biggr]^2,
  \end{split}
\end{align}
where, $t$ and $t^\prime$ are the compared times. The summation loops over all $N$ particles $j$ in the vicinity of particle $i$, and $\vec{\epsilon}$ is the uniform strain of the region. 
The minimum $D_\text{min}=\min_{\vec{\epsilon}}|D|$ is obtained by finding an appropriate strain $\vec{\epsilon}$ and quantifies the local deviation from the affine deformation. A high value of $D_\text{min}$ indicates small-scale plastic events (mechanism~1), while a low value indicates movement in the bulk (mechanism~2). In the following, the particles' $D_\text{min}$ are obtained by comparing two snapshots $\SI{2.5}{\second}$ apart. \red{Particles with less than 3 contacts are excluded from the analysis.}

\begin{figure}[htb!]
  \centering
\includegraphics[width=\linewidth]{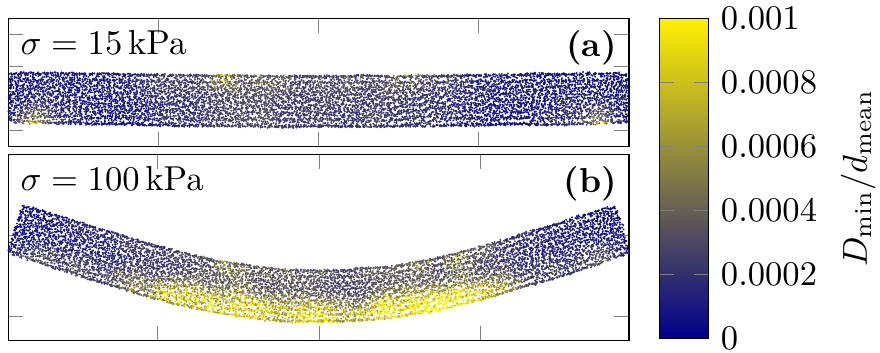}
  \caption{$D_\text{min}/d_\text{mean}$ of the beam. Examples are displayed for systems with the particles' stiffness $E_\text{p}=\SI{100}{\mega\pascal}$ and friction coefficient $\mu_\text{p}=0.3$ and external stress values of (a)~$\sigma=\SI{15}{\kilo\pascal}$ and (b)~$\sigma=\SI{100}{\kilo\pascal}$.}
  \label{fig:d2MinSnapshot}
\end{figure}

Examples of $D_\text{min}/d_\text{mean}$ values within the beam are shown in Fig.~\ref{fig:d2MinSnapshot} for two different stages of bending. Regions of localized and bulk deformation are shown respectively in \red{yellow (clear) and blue (dark).} 
Note that the value of $D_\text{min}/d_\text{mean}$ remains small throughout the experiment, showing that our jammed phase tends to deform in an affine motion in general.

\begin{figure}[htb!]
  \centering
%  \printfig{ \input{media/img/d2MinAutoCorr/d2AutoCorrHorizontalProtB}}
    % \input{media/img/d2MinAutoCorr/d2TimeAutoCorrHorizontalProtB}
\includegraphics{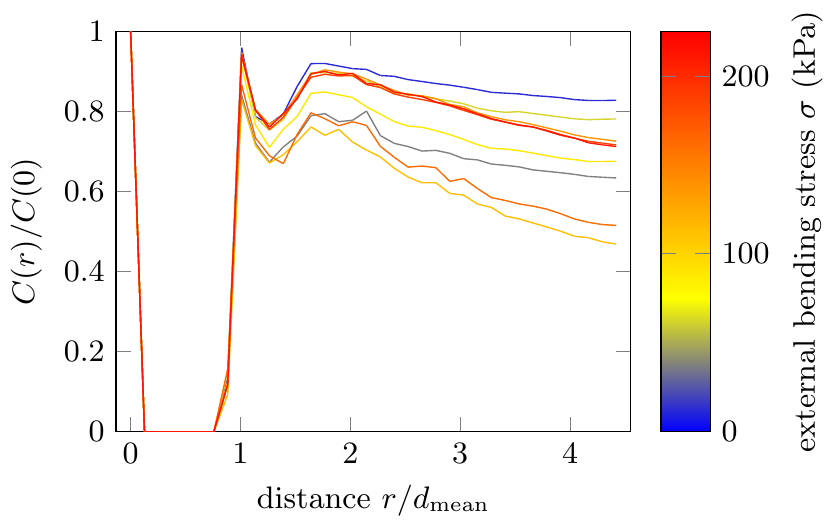}
  \vspace{-5mm}
  \caption{\red{Spatial} auto-correlation function of $D_\text{min}$. Different lines are averaged over $\sigma$ for intervals of $\SI{25}{\kilo\pascal}$. The line color indicates the intervals' average value. The displayed data is obtained for $\mu_\text{p}=0.3$ and $E_\text{p}=\SI{100}{\mega\pascal}$ with protocol \textsc{b}.}
  \label{fig:d2AutoCorr}
\end{figure}

The spatial auto-correlation function quantifies the length-scale on which particles undergo a similar degree of localized deformation. Its definition is given in Appendix~\ref{supp:autoCorr}.
Figure~\ref{fig:d2AutoCorr} shows the auto-correlation, $C(r)$, for different stages of bending stress throughout the simulation. 
\red{The correlation of neighboring particles is high, $C(r)/C(0)>0.8$. Although it decreases for an increasing distance $r$, it remains high within the investigated region. This indicates that all particles deviate from the affine transformation in a similar manner. There is no systematic difference between different stages of bending.}
% \todo{Look at the following}
% The temporal auto-correlation function (inset of Figure~\ref{fig:d2AutoCorr}) shows a high correlation for large time differences.
% Combined with the limited spatial correlation (described above), this confirms that the location of regions of high local deformation persists through time.

\subsection{Local density variations}
\label{sec:res:loc-volume-fraction}

\begin{figure}[hbt]
  \centering
\includegraphics[width=\linewidth]{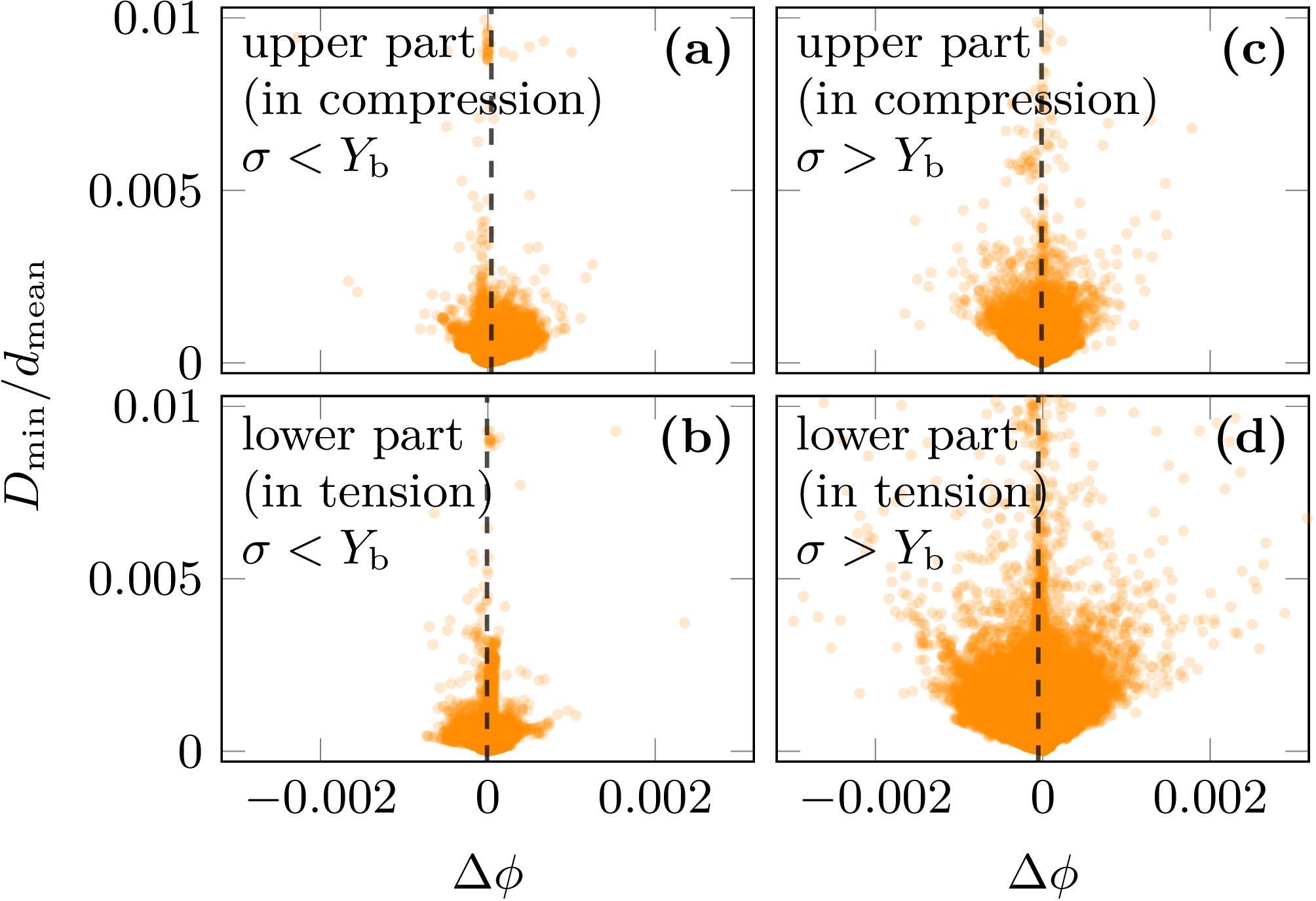}
  \vspace{-5mm}
  \caption{$D_\text{min}$ value of each particle divided by the mean particle diameter $d_\text{mean}$, as a function of the change in each particle's local volume fraction, $\Delta\phi$, obtained by Voronoi tessellation. The beam is divided into: (a, c) its top part, in compression, and (b, d) its bottom part, in tension. A particle is assumed to be in the top part if its distance to the top membrane is shorter than its distance to the bottom membrane. Otherwise, it is assumed to be within the bottom part of the beam.
  Plots (a, b) display data for $\sigma<Y_\text{b}$, plots (c, d)  for $\sigma>Y_\text{b}$. Points are plotted semitransparent to visualize regions with many overlapping points. Dashed vertical lines indicate the average $\langle \Delta\phi \rangle$ for all points in each subfigure. All plots are obtained for $\mu_\text{p}=0.3$, $E=\SI{100}{\mega\pascal}$, simulated with protocol~\textsc{a}.}
  \label{fig:d2MinVolumeFractionScatter}
\end{figure}

To quantify the change in local volume fraction associated with the deformation mode, 
we plot $D_\text{min}/d_\text{mean}$ of every particle in the system, 
against each particle's change in local volume fraction, $\Delta\phi$, in Figure~\ref{fig:d2MinVolumeFractionScatter}. For this analysis, we calculate a particle's local volume fraction by dividing the particle's volume by the volume of its Voronoi cell.
Dashed lines represent the average change in local volume fraction, $\langle \Delta\phi \rangle$.
\red{For a low stress $\sigma<Y_\text{b}$, on average, the local volume fraction of particles in the top part of the beam (Fig.~\ref{fig:d2MinVolumeFractionScatter}a) increases ($\langle \Delta\phi \rangle > 0$) due to compression.
For high stress $\sigma>Y_\text{b}$, the local volume fraction decreases on average ($\langle \Delta\phi \rangle < 0$) in the top part of the beam. In the bottom part of the beam the local volume fraction decreases on average, independent of the external stress $\sigma$. The absolute value of $\Delta\phi$ of the individual particles increases with increasing stress, and so does the $D_\text{min}$ values, becoming greater for high bending stresses.}

\red{The deviation from the affine deformation, $D_\text{min}$, is correlated with the change in local particle volume fraction:
a large $D_\text{min}$ corresponds to a large $\Delta\phi$. This is visible from the \textsc{v}-shaped lower edge of all points in Fig.~\ref{fig:d2MinVolumeFractionScatter}.}

\subsection{Spatial distribution of local deformation}

\begin{figure}[h]
%    \printfig{  \includegraphics[width=\linewidth, axisratio=1.3]{media/img/d2Min/d2PerHeightE1e8Mu0.3.tikz}}
\includegraphics{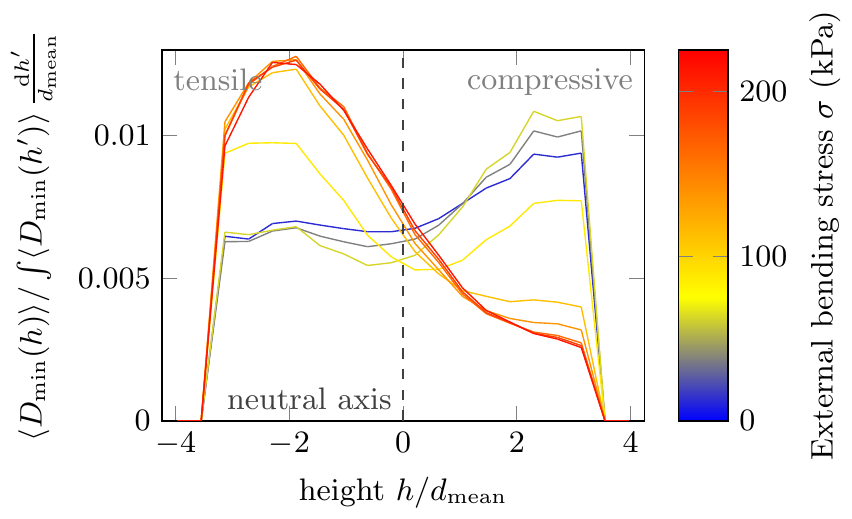}
\vspace{-5mm}
    \caption{Normalized $D_\text{min}$ as a function of height, $h$. The height is measured from the center plane of the beam, which is the geometrical midplane between the beam's top and bottom surface. Negative values indicate positions in the \red{tensile} part of the beam, while positive values indicate positions in the \red{compressive} part of the beam. Different lines are averaged over $\sigma$ for intervals of $\SI{25}{\kilo\pascal}$. The line color indicates the intervals mean value. The displayed data is obtained for $\mu_\text{p}=0.3$ and $E_\text{p}=\SI{100}{\mega\pascal}$.}
    \label{fig:d2MinHeightStress}
\end{figure}

The data displayed in Figure~\ref{fig:d2MinVolumeFractionScatter} suggests
that there is a difference in $D_\text{min}$ between the beam's top part (in compression) and its bottom part (in tension).
Therefore, we look at the distribution of $D_\text{min}$ as a function of height, $h$, in the beam. Figure~\ref{fig:d2MinHeightStress} shows the distribution for different intervals of stress, $\sigma$, throughout the bending process. 
\red{In the initial stage of bending, $D_\text{min}$ is larger in the top (compressive) part of the beam than in its bottom part.} For high $\sigma$, $D_\text{min}$ is larger in the lower part of the beam (in tension). \red{The transition between those behaviors happens around the beam's ultimate strength at $\sigma\approx Y_\text{b}$.}

\begin{figure}[h]
%  \printfig{\includegraphics[width=\linewidth, axisratio=1.6]{media/img/d2Min/d2PerHeightE1e8Sigma175_200.tikz}}
\includegraphics{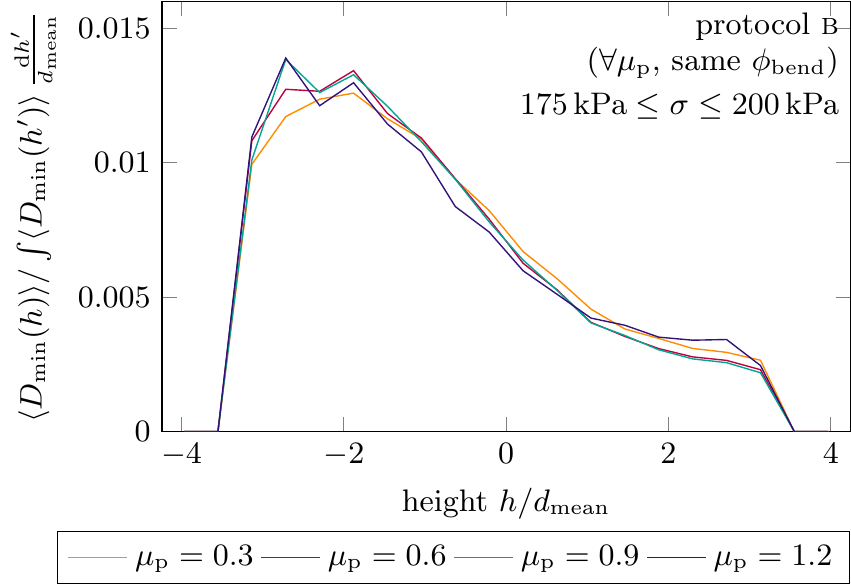}
  \caption{Normalized $D_\text{min}$ as a function of height, $h$, for friction coefficients $\mu_\text{p}\in\{0.3,\,0.6,\,0.9,\,1.2\}$. The height is measured from the center plane of the beam, which is the geometrical midplane between the beam's top and bottom surface. Negative values indicate positions in the \red{tensile} part of the beam, while positive values indicate positions in the \red{compressive} part of the beam. The plot is obtained for $\SI{175}{\kilo\pascal}\leq\sigma\leq\SI{200}{\kilo\pascal}$ with protocol \textsc{b}.}
  \label{fig:d2MinHeightStressMu}
\end{figure}
  
\begin{figure*}[!hbt]  
  \centering
%\printfig{  \begin{tikzpicture}[font=\small]
%    \input{media/img/volumeFractionD2/d2MinAvgVolumeFraction_E10.tex}
%    \input{media/img/volumeFractionD2/d2MinAvgVolumeFraction_E100.tex}
%  \end{tikzpicture}
%  }
\includegraphics{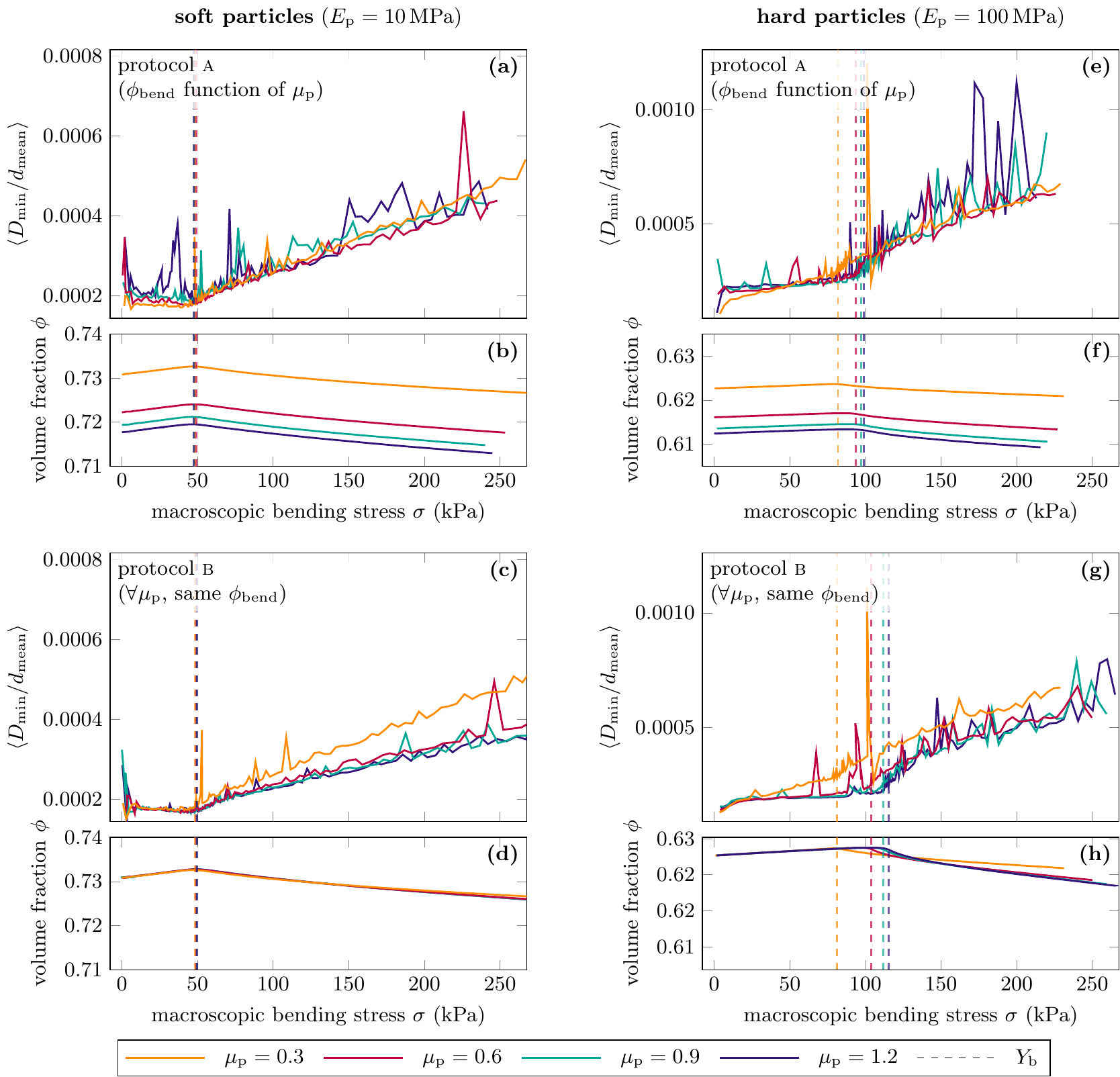}
  \caption{The average $D_\text{min}$ value divided by the mean particle diameter $d_\text{mean}$ (a, c, e, g) and the volume fraction $\phi$ (b, d, f, h) as a function of applied bending stress. Simulations are executed with protocol \textsc{a} for plots (a, b, e, f) and protocol \textsc{b} for plots (c, d, g, h). Vertical lines drawn at $\sigma=Y_\text{b}(\mu_\text{p})$ indicate the ultimate stress obtained from the stress-strain measurement. Please note the different axis scales between the left and right panels (respectively $E_\text{p}=\SI{10}{\mega\pascal}$ and $E_\text{p}=\SI{100}{\mega\pascal}$).}
  \label{fig:d2MinVolumeFraction}
\end{figure*}

Figure~\ref{fig:d2MinHeightStressMu} compares $D_\text{min}$ along the beam's height for different friction coefficients at high $\sigma$, towards the end of the bending experiment. 
For all friction coefficients, we observe a similar spatial distribution. However, high particles' friction leads to a more pronounced difference between the top and bottom parts of the beam, where regions of non-affine deformation appear shifted towards the bottom part of the beam, in tension. 
In other words, high friction inhibits motion where the granulate is in compression, 
hence less plastic events happen in the top part of the beam. 
Note that there is no significant difference between the two protocols, although normalized $D_\text{min}$ values obtained with protocol \textsc{a} (not shown here) show closer values, \red{because the effect of friction is weakened by the differing volume fractions $\phi_\text{bend}$ at the onset of bending}.

\subsection{Temporal distribution of local plastic events}

The spatial average $\langle D_\text{min}\rangle$
as a function of macroscopic bending stress, $\sigma$,
can further clarify how the beam deforms throughout the bending process. 
Figure~\ref{fig:d2MinVolumeFraction} displays the evolution of $\langle D_\text{min}\rangle$ and the corresponding volume fraction $\phi$, throughout the bending test, for different parameter combinations.

As observed for the average force per contact, $\langle D_\text{min}\rangle$ is non-monotonic, dividing each bending test in two regions around a macroscopic stress approximately equal to the beam's ultimate strength, $\sigma\approx Y_\text{b}$. For low stress, $\sigma<Y_\text{b}$, $\langle D_\text{min}\rangle$ decreases with increasing stress. During this initial stage of bending, the system densifies. 
Affine contraction of the granular fabric
allows the volume fraction to increase, which leads to less individual particle motion and less localized deformations (decreasing $\langle D_\text{min}\rangle$ value).

During the second stage of bending, i.e.~for high stress, $\sigma>Y_\text{b}$, $\langle D_\text{min}\rangle$ increases with increasing stress. Concurrently, the volume fraction, $\phi$, decreases, effectively leaving more space for the particles to move. 
This is true specifically in the beam's bottom part, below the neutral axis, where the beam is in tension.
This second regime corresponds to the destabilization of the force network observed in Sec.~\ref{sec:forceNetworkRobustness.},
which happens simultaneously with that of the contact network.

Let us look at how the particles' properties influence the system's behavior. Although all systems, independently of the particles' elastic modulus or friction coefficient, exhibit the two regimes described above, both quantities play a role. 

Looking at $\langle D_\text{min}\rangle$ for the particles' elastic modulus, we find that it differs by almost one order of magnitude between the system with soft particles ($E_\text{p}=\SI{10}{\mega\pascal}$, Fig.~\ref{fig:d2MinVolumeFraction}a-d) and hard particles ($E_\text{p}=\SI{100}{\mega\pascal}$, Fig.~\ref{fig:d2MinVolumeFraction}e-h). This is a result of the denser system obtained for soft particles: the higher deformability of the soft particles allows them to achieve a higher volume fraction than hard particles, at the same confining pressure. A denser packing hampers relative motion of the particles, and thereby localized non-affine deformations.

Focusing further on the influence of the particles' friction, we find that for protocol \textsc{b}, i.e. systems where the initial volume fraction $\phi_\text{bend}$ is independent of $\mu_\text{p}$ (Fig.~\ref{fig:d2MinVolumeFraction}c, d, g, h), the volume fraction $\phi$ and $\langle D_\text{min}\rangle$ have a low spread and are similar for all friction coefficients. 
\red{Only for a friction coefficient of $\mu_\text{p}=0.3$, $\langle D_\text{min}\rangle$ separates from the ones obtained for other friction coefficients (Fig.~\ref{fig:d2MinVolumeFraction}c, g)}. The separation occurs because particles with lower friction are able to move under a lower external load in a system of equal density. For $\mu_\text{p}>0.3$, $\langle D_\text{min}\rangle$ are also expected to deviate one by one from that of larger $\mu_\text{p}$, as the stress increases beyond the studied region and allows more contacts to overcome the Coulomb friction threshold -- and that, for gradually higher $\mu_\text{p}$. Systems simulated with protocol \textsc{a}, i.e. systems where the initial volume fraction $\phi_\text{bend}$ is a function of $\mu_\text{p}$, exhibit a larger spread in volume fraction $\phi$ and localized deformations $\langle D_\text{min}\rangle$, compared to systems simulated with protocol \textsc{b}. Systems starting with a lower volume fraction (i.e.~higher $\mu_\text{p}$) provide more space to each individual particle and hence allow for more local deformation.

\section{Conclusion}
The macroscopic properties of granular metamaterials depend heavily on the properties of their constituent particles, and on the spatial arrangement of these constituents. In this work, we analyzed the influence of the particles' elastic modulus and friction coefficient on the mechanical properties of a jammed granular beam. We particularly studied the mesoscopic-scale effect of particles' properties by investigating the mechanisms through which these influence the beam's macroscopic properties in the case of jammed granular media.

We find that higher particles' elastic modulus increases the stiffness and ultimate strength of the beam, although hard particles pack less densely than soft particles, the latter being able to deform into a very dense jammed phase.

Particles' friction $\mu_\text{p}$ has a more complex influence on the jammed-granulate macroscopic response.
Friction is necessary for the stability of jammed packings.
Without friction ($\mu_\text{p}=0$), there is no resistance to particles sliding against each other, 
and minimal loading can cause large-scale reorganization.  
If the preparation of the packing is done at the same particles' friction coefficient, $\mu_\text{p}$, as the bending experiment, high $\mu_\text{p}$ increase the beam's ultimate strength, but contrary to intuition, decrease the beam's stiffness. By varying initial conditions, we showed that this decrease in the beam's stiffness is an effect of volume fraction: 
for beams composed of a packing of equal volume fraction, the particle's friction does not have influence on the macroscopic stiffness of the jammed phase. 
The ultimate strength $Y_\text{b}(\mu_\text{p})$, however, is a function of the friction coefficient.

These results highlight that both particles' material properties and preparation protocol are essential in determining the macroscopic properties of jammed-granulate metamaterials.

Beyond stiffness and ultimate strength, we analyzed the internal structure of the granular beam: on isotropically jammed granular phases, we showed that the ultimate strength measured from stress-strain curves coincides with a failure of the force network. This failure constitutes a delimitation between two regimes, persistent for all particles' properties studied. Prior to failure, each contact carries an increasing amount of load. From the force network failure and beyond, contacts break to release stress, while an increasing amount of contacts gets close to the Coulomb threshold.

We further demonstrated that deformation happens 
in a generally affine manner, meaning that
the granular network is compressed or extended in a coordinated motion of particles.
When local plastic events are observed, they happen for clusters of approx.~four particles in diameter behaving similarly.
Those clusters of motion are 
spread homogeneously throughout the beam, and persist throughout bending, both in size and in location.
Before failure of the force network, those events mostly contribute to densification in the compressive part of the beam (top half), while they occur due to lower volume fractions in the tensile part (bottom half) for higher stresses.

Our findings provide guidelines for selecting granular particles and a suited preparation protocol, towards the creation of jammed-granulate metamaterials with targeted, application-specific mechanical properties. To maximize the stiffness, particles with a large elastic modulus should be preferred, and a preparation protocol resulting in high packing density should be implemented. To maximize the ultimate strength, particles with high friction should be chosen, while still keeping the packing dense. 
\red{Tunability is maximized by varying $\mu_\text{p}$ from very small, $\mu_\text{p} \to 0$, to intermediate, $\mu_\text{p} \approx 0.6$.}

\red{
We also showed that granular packing density, $\phi$, can be used as a tuning parameter in such system. Although protocol~\textsc{b} might seem artificial from an experimental perspective, its goal, namely to create an initial packing at a density independent of the particles' friction coefficient, can be obtained experimentally:
\begin{enumerate}
  \item Packing density can be increased by pre-constraining the granulate once it is enclosed by the membrane during manufacturing. This can be achieved by imposing reorganizations by deformation while maintaining pressure to force the granular media into a denser configuration.
  \item Vibrating the system, during or after manufacturing, will increase the packing density.
  \item By using particulate media whose friction, particle size, or stiffness can be altered~\cite{Perrin2019, Rudge2020}, for example by temperature, charge, or water content, packing density can all be tweaked.
\end{enumerate}}
Further exploration, notably experimental, of collective effects within a jammed medium under load, will deepen our understanding of amorphous solids and inform novel concepts for granular metamaterials. 

\bibliography{jammed-beam-micro-macro.bib}

\begin{thebibliography}{54}%
\makeatletter
\providecommand \@ifxundefined [1]{%
 \@ifx{#1\undefined}
}%
\providecommand \@ifnum [1]{%
 \ifnum #1\expandafter \@firstoftwo
 \else \expandafter \@secondoftwo
 \fi
}%
\providecommand \@ifx [1]{%
 \ifx #1\expandafter \@firstoftwo
 \else \expandafter \@secondoftwo
 \fi
}%
\providecommand \natexlab [1]{#1}%
\providecommand \enquote  [1]{``#1''}%
\providecommand \bibnamefont  [1]{#1}%
\providecommand \bibfnamefont [1]{#1}%
\providecommand \citenamefont [1]{#1}%
\providecommand \href@noop [0]{\@secondoftwo}%
\providecommand \href [0]{\begingroup \@sanitize@url \@href}%
\providecommand \@href[1]{\@@startlink{#1}\@@href}%
\providecommand \@@href[1]{\endgroup#1\@@endlink}%
\providecommand \@sanitize@url [0]{\catcode `\\12\catcode `\$12\catcode `\&12\catcode `\#12\catcode `\^12\catcode `\_12\catcode `\%12\relax}%
\providecommand \@@startlink[1]{}%
\providecommand \@@endlink[0]{}%
\providecommand \url  [0]{\begingroup\@sanitize@url \@url }%
\providecommand \@url [1]{\endgroup\@href {#1}{\urlprefix }}%
\providecommand \urlprefix  [0]{URL }%
\providecommand \Eprint [0]{\href }%
\providecommand \doibase [0]{https://doi.org/}%
\providecommand \selectlanguage [0]{\@gobble}%
\providecommand \bibinfo  [0]{\@secondoftwo}%
\providecommand \bibfield  [0]{\@secondoftwo}%
\providecommand \translation [1]{[#1]}%
\providecommand \BibitemOpen [0]{}%
\providecommand \bibitemStop [0]{}%
\providecommand \bibitemNoStop [0]{.\EOS\space}%
\providecommand \EOS [0]{\spacefactor3000\relax}%
\providecommand \BibitemShut  [1]{\csname bibitem#1\endcsname}%
\let\auto@bib@innerbib\@empty
%</preamble>
\bibitem [{\citenamefont {Liu}\ and\ \citenamefont {Nagel}(2001)}]{Liu2001}%
  \BibitemOpen
  \bibinfo {editor} {\bibfnamefont {A.~J.}\ \bibnamefont {Liu}}\ and\ \bibinfo {editor} {\bibfnamefont {S.~R.}\ \bibnamefont {Nagel}},\ eds.,\ \href@noop {} {\emph {\bibinfo {title} {Jamming and Rheology: Constrained Dynamics on Microscopic and Macroscopic Scales}}}\ (\bibinfo  {publisher} {Taylor {\&} Francis},\ \bibinfo {address} {New York},\ \bibinfo {year} {2001})\BibitemShut {NoStop}%
\bibitem [{\citenamefont {O'Hern}\ \emph {et~al.}(2003)\citenamefont {O'Hern}, \citenamefont {Silbert}, \citenamefont {Liu},\ and\ \citenamefont {Nagel}}]{OHern2003}%
  \BibitemOpen
  \bibfield  {author} {\bibinfo {author} {\bibfnamefont {C.~S.}\ \bibnamefont {O'Hern}}, \bibinfo {author} {\bibfnamefont {L.~E.}\ \bibnamefont {Silbert}}, \bibinfo {author} {\bibfnamefont {A.~J.}\ \bibnamefont {Liu}},\ and\ \bibinfo {author} {\bibfnamefont {S.~R.}\ \bibnamefont {Nagel}},\ }\bibfield  {title} {\bibinfo {title} {Jamming at zero temperature and zero applied stress: The epitome of disorder},\ }\href {https://doi.org/10.1103/PhysRevE.68.011306} {\bibfield  {journal} {\bibinfo  {journal} {Phys. Rev. E}\ }\textbf {\bibinfo {volume} {68}},\ \bibinfo {pages} {011306} (\bibinfo {year} {2003})}\BibitemShut {NoStop}%
\bibitem [{\citenamefont {Fitzgerald}\ \emph {et~al.}(2020)\citenamefont {Fitzgerald}, \citenamefont {Delaney},\ and\ \citenamefont {Howard}}]{Fitzgerald2020}%
  \BibitemOpen
  \bibfield  {author} {\bibinfo {author} {\bibfnamefont {S.~G.}\ \bibnamefont {Fitzgerald}}, \bibinfo {author} {\bibfnamefont {G.~W.}\ \bibnamefont {Delaney}},\ and\ \bibinfo {author} {\bibfnamefont {D.}~\bibnamefont {Howard}},\ }\bibfield  {title} {\bibinfo {title} {A review of jamming actuation in soft robotics},\ }\href {https://doi.org/10.3390/act9040104} {\bibfield  {journal} {\bibinfo  {journal} {Actuators}\ }\textbf {\bibinfo {volume} {9}},\ \bibinfo {pages} {104} (\bibinfo {year} {2020})}\BibitemShut {NoStop}%
\bibitem [{\citenamefont {Pendry}(2000)}]{pendryNegativeRefractionMakes2000}%
  \BibitemOpen
  \bibfield  {author} {\bibinfo {author} {\bibfnamefont {J.~B.}\ \bibnamefont {Pendry}},\ }\bibfield  {title} {\bibinfo {title} {Negative {{Refraction Makes}} a {{Perfect Lens}}},\ }\href {https://doi.org/10.1103/PhysRevLett.85.3966} {\bibfield  {journal} {\bibinfo  {journal} {Phys. Rev. Lett.}\ }\textbf {\bibinfo {volume} {85}},\ \bibinfo {pages} {3966} (\bibinfo {year} {2000})}\BibitemShut {NoStop}%
\bibitem [{\citenamefont {Loeve}\ \emph {et~al.}(2010)\citenamefont {Loeve}, \citenamefont {van~de Ven}, \citenamefont {Vogel}, \citenamefont {Breedveld},\ and\ \citenamefont {Dankelman}}]{Loeve2010}%
  \BibitemOpen
  \bibfield  {author} {\bibinfo {author} {\bibfnamefont {A.~J.}\ \bibnamefont {Loeve}}, \bibinfo {author} {\bibfnamefont {O.~S.}\ \bibnamefont {van~de Ven}}, \bibinfo {author} {\bibfnamefont {J.~G.}\ \bibnamefont {Vogel}}, \bibinfo {author} {\bibfnamefont {P.}~\bibnamefont {Breedveld}},\ and\ \bibinfo {author} {\bibfnamefont {J.}~\bibnamefont {Dankelman}},\ }\bibfield  {title} {\bibinfo {title} {Vacuum packed particles as flexible endoscope guides with controllable rigidity},\ }\href {https://doi.org/10.1007/s10035-010-0193-8} {\bibfield  {journal} {\bibinfo  {journal} {Granular Matter}\ }\textbf {\bibinfo {volume} {12}},\ \bibinfo {pages} {543} (\bibinfo {year} {2010})}\BibitemShut {NoStop}%
\bibitem [{\citenamefont {Jiang}\ \emph {et~al.}(2014)\citenamefont {Jiang}, \citenamefont {Dasgupta}, \citenamefont {Althoefer},\ and\ \citenamefont {Nanayakkara}}]{jiangRoboticGranularJamming2014a}%
  \BibitemOpen
  \bibfield  {author} {\bibinfo {author} {\bibfnamefont {A.}~\bibnamefont {Jiang}}, \bibinfo {author} {\bibfnamefont {P.}~\bibnamefont {Dasgupta}}, \bibinfo {author} {\bibfnamefont {K.}~\bibnamefont {Althoefer}},\ and\ \bibinfo {author} {\bibfnamefont {T.}~\bibnamefont {Nanayakkara}},\ }\bibfield  {title} {\bibinfo {title} {Robotic granular jamming: A new variable stiffness mechanism},\ }\href {https://doi.org/10.7210/jrsj.32.333} {\bibfield  {journal} {\bibinfo  {journal} {JRSJ}\ }\textbf {\bibinfo {volume} {32}},\ \bibinfo {pages} {333} (\bibinfo {year} {2014})}\BibitemShut {NoStop}%
\bibitem [{\citenamefont {G{\"o}tz}\ and\ \citenamefont {P{\"o}schel}(2023{\natexlab{a}})}]{gotzGranularMetaMaterialViscoelastic2022}%
  \BibitemOpen
  \bibfield  {author} {\bibinfo {author} {\bibfnamefont {H.}~\bibnamefont {G{\"o}tz}}\ and\ \bibinfo {author} {\bibfnamefont {T.}~\bibnamefont {P{\"o}schel}},\ }\bibfield  {title} {\bibinfo {title} {Granular meta-material: Response of a bending beam},\ }\href {https://doi.org/10.1007/s10035-023-01336-9} {\bibfield  {journal} {\bibinfo  {journal} {Granular Matter}\ }\textbf {\bibinfo {volume} {25}},\ \bibinfo {pages} {58} (\bibinfo {year} {2023}{\natexlab{a}})}\BibitemShut {NoStop}%
\bibitem [{\citenamefont {Chaudhuri}\ \emph {et~al.}(2010)\citenamefont {Chaudhuri}, \citenamefont {Berthier},\ and\ \citenamefont {Sastry}}]{Chaudhuri2010}%
  \BibitemOpen
  \bibfield  {author} {\bibinfo {author} {\bibfnamefont {P.}~\bibnamefont {Chaudhuri}}, \bibinfo {author} {\bibfnamefont {L.}~\bibnamefont {Berthier}},\ and\ \bibinfo {author} {\bibfnamefont {S.}~\bibnamefont {Sastry}},\ }\bibfield  {title} {\bibinfo {title} {Jamming transitions in amorphous packings of frictionless spheres occur over a continuous range of volume fractions},\ }\href {https://doi.org/10.1103/PhysRevLett.104.165701} {\bibfield  {journal} {\bibinfo  {journal} {Physical Review Letters}\ }\textbf {\bibinfo {volume} {104}},\ \bibinfo {pages} {165701} (\bibinfo {year} {2010})}\BibitemShut {NoStop}%
\bibitem [{\citenamefont {Pica~Ciamarra}\ \emph {et~al.}(2010)\citenamefont {Pica~Ciamarra}, \citenamefont {Coniglio},\ and\ \citenamefont {de~Candia}}]{Ciamarra2010}%
  \BibitemOpen
  \bibfield  {author} {\bibinfo {author} {\bibfnamefont {M.}~\bibnamefont {Pica~Ciamarra}}, \bibinfo {author} {\bibfnamefont {A.}~\bibnamefont {Coniglio}},\ and\ \bibinfo {author} {\bibfnamefont {A.}~\bibnamefont {de~Candia}},\ }\bibfield  {title} {\bibinfo {title} {Disordered jammed packings of frictionless spheres},\ }\href {https://doi.org/10.1039/C001904F} {\bibfield  {journal} {\bibinfo  {journal} {Soft Matter}\ }\textbf {\bibinfo {volume} {6}},\ \bibinfo {pages} {13} (\bibinfo {year} {2010})}\BibitemShut {NoStop}%
\bibitem [{\citenamefont {Otsuki}\ and\ \citenamefont {Hayakawa}(2012)}]{Otsuki2012}%
  \BibitemOpen
  \bibfield  {author} {\bibinfo {author} {\bibfnamefont {M.}~\bibnamefont {Otsuki}}\ and\ \bibinfo {author} {\bibfnamefont {H.}~\bibnamefont {Hayakawa}},\ }\bibfield  {title} {\bibinfo {title} {Critical scaling of a jammed system after a quench of temperature},\ }\href {https://doi.org/10.1103/PhysRevE.86.031505} {\bibfield  {journal} {\bibinfo  {journal} {Physical Review E}\ }\textbf {\bibinfo {volume} {86}},\ \bibinfo {pages} {031505} (\bibinfo {year} {2012})}\BibitemShut {NoStop}%
\bibitem [{\citenamefont {Hermes}\ and\ \citenamefont {Dijkstra}(2010)}]{Hermes2010}%
  \BibitemOpen
  \bibfield  {author} {\bibinfo {author} {\bibfnamefont {M.}~\bibnamefont {Hermes}}\ and\ \bibinfo {author} {\bibfnamefont {M.}~\bibnamefont {Dijkstra}},\ }\bibfield  {title} {\bibinfo {title} {Jamming of polydisperse hard spheres: the effect of kinetic arrest},\ }\href {https://doi.org/10.1209/0295-5075/89/38005} {\bibfield  {journal} {\bibinfo  {journal} {{EPL} (Europhysics Letters)}\ }\textbf {\bibinfo {volume} {89}},\ \bibinfo {pages} {38005} (\bibinfo {year} {2010})}\BibitemShut {NoStop}%
\bibitem [{\citenamefont {Baranau}\ and\ \citenamefont {Tallarek}(2014)}]{Baranau2014}%
  \BibitemOpen
  \bibfield  {author} {\bibinfo {author} {\bibfnamefont {V.}~\bibnamefont {Baranau}}\ and\ \bibinfo {author} {\bibfnamefont {U.}~\bibnamefont {Tallarek}},\ }\bibfield  {title} {\bibinfo {title} {Random-close packing limits for monodisperse and polydisperse hard spheres},\ }\href {https://doi.org/10.1039/C3SM52959B} {\bibfield  {journal} {\bibinfo  {journal} {Soft Matter}\ }\textbf {\bibinfo {volume} {10}},\ \bibinfo {pages} {21} (\bibinfo {year} {2014})}\BibitemShut {NoStop}%
\bibitem [{\citenamefont {Torquato}\ and\ \citenamefont {Stillinger}(2010)}]{Torquato2010}%
  \BibitemOpen
  \bibfield  {author} {\bibinfo {author} {\bibfnamefont {S.}~\bibnamefont {Torquato}}\ and\ \bibinfo {author} {\bibfnamefont {F.~H.}\ \bibnamefont {Stillinger}},\ }\bibfield  {title} {\bibinfo {title} {Jammed hard-particle packings: From {K}epler to {B}ernal and beyond},\ }\href {https://doi.org/10.1103/RevModPhys.82.2633} {\bibfield  {journal} {\bibinfo  {journal} {Rev. Mod. Phys.}\ }\textbf {\bibinfo {volume} {82}},\ \bibinfo {pages} {2633} (\bibinfo {year} {2010})}\BibitemShut {NoStop}%
\bibitem [{\citenamefont {Kumar}\ and\ \citenamefont {Luding}(2016)}]{Kumar2016}%
  \BibitemOpen
  \bibfield  {author} {\bibinfo {author} {\bibfnamefont {N.}~\bibnamefont {Kumar}}\ and\ \bibinfo {author} {\bibfnamefont {S.}~\bibnamefont {Luding}},\ }\bibfield  {title} {\bibinfo {title} {Memory of jamming--multiscale models for soft and granular matter},\ }\href {https://doi.org/10.1007/s10035-016-0624-2} {\bibfield  {journal} {\bibinfo  {journal} {Granular Matter}\ }\textbf {\bibinfo {volume} {18}},\ \bibinfo {pages} {3} (\bibinfo {year} {2016})}\BibitemShut {NoStop}%
\bibitem [{\citenamefont {D'Angelo}\ \emph {et~al.}(2022)\citenamefont {D'Angelo}, \citenamefont {Horb}, \citenamefont {Cowley}, \citenamefont {Sperl},\ and\ \citenamefont {Kranz}}]{DAngelo2022}%
  \BibitemOpen
  \bibfield  {author} {\bibinfo {author} {\bibfnamefont {O.}~\bibnamefont {D'Angelo}}, \bibinfo {author} {\bibfnamefont {A.}~\bibnamefont {Horb}}, \bibinfo {author} {\bibfnamefont {A.}~\bibnamefont {Cowley}}, \bibinfo {author} {\bibfnamefont {M.}~\bibnamefont {Sperl}},\ and\ \bibinfo {author} {\bibfnamefont {W.~T.}\ \bibnamefont {Kranz}},\ }\bibfield  {title} {\bibinfo {title} {Granular piston-probing in microgravity: powder compression, from densification to jamming},\ }\href {https://doi.org/10.1038/s41526-022-00235-2} {\bibfield  {journal} {\bibinfo  {journal} {npj Microgravity}\ }\textbf {\bibinfo {volume} {8}},\ \bibinfo {pages} {48} (\bibinfo {year} {2022})}\BibitemShut {NoStop}%
\bibitem [{\citenamefont {van Hecke}(2009)}]{VanHecke2009}%
  \BibitemOpen
  \bibfield  {author} {\bibinfo {author} {\bibfnamefont {M.}~\bibnamefont {van Hecke}},\ }\bibfield  {title} {\bibinfo {title} {Jamming of soft particles: geometry, mechanics, scaling and isostaticity},\ }\href {https://doi.org/10.1088/0953-8984/22/3/033101} {\bibfield  {journal} {\bibinfo  {journal} {Journal of Physics: Condensed Matter}\ }\textbf {\bibinfo {volume} {22}},\ \bibinfo {pages} {033101} (\bibinfo {year} {2009})}\BibitemShut {NoStop}%
\bibitem [{\citenamefont {Ciamarra}\ \emph {et~al.}(2011)\citenamefont {Ciamarra}, \citenamefont {Pastore}, \citenamefont {Nicodemi},\ and\ \citenamefont {Coniglio}}]{Ciamarra2011}%
  \BibitemOpen
  \bibfield  {author} {\bibinfo {author} {\bibfnamefont {M.~P.}\ \bibnamefont {Ciamarra}}, \bibinfo {author} {\bibfnamefont {R.}~\bibnamefont {Pastore}}, \bibinfo {author} {\bibfnamefont {M.}~\bibnamefont {Nicodemi}},\ and\ \bibinfo {author} {\bibfnamefont {A.}~\bibnamefont {Coniglio}},\ }\bibfield  {title} {\bibinfo {title} {Jamming phase diagram for frictional particles},\ }\href {https://doi.org/10.1103/PhysRevE.84.041308} {\bibfield  {journal} {\bibinfo  {journal} {Physical Review E}\ }\textbf {\bibinfo {volume} {84}},\ \bibinfo {pages} {041308} (\bibinfo {year} {2011})}\BibitemShut {NoStop}%
\bibitem [{\citenamefont {Mari}\ \emph {et~al.}(2014)\citenamefont {Mari}, \citenamefont {Seto}, \citenamefont {Morris},\ and\ \citenamefont {Denn}}]{Mari2014}%
  \BibitemOpen
  \bibfield  {author} {\bibinfo {author} {\bibfnamefont {R.}~\bibnamefont {Mari}}, \bibinfo {author} {\bibfnamefont {R.}~\bibnamefont {Seto}}, \bibinfo {author} {\bibfnamefont {J.~F.}\ \bibnamefont {Morris}},\ and\ \bibinfo {author} {\bibfnamefont {M.~M.}\ \bibnamefont {Denn}},\ }\bibfield  {title} {\bibinfo {title} {{Shear thickening, frictionless and frictional rheologies in non-Brownian suspensions}},\ }\href {https://doi.org/10.1122/1.4890747} {\bibfield  {journal} {\bibinfo  {journal} {Journal of Rheology}\ }\textbf {\bibinfo {volume} {58}},\ \bibinfo {pages} {1693} (\bibinfo {year} {2014})}\BibitemShut {NoStop}%
\bibitem [{\citenamefont {Bi}\ \emph {et~al.}(2011)\citenamefont {Bi}, \citenamefont {Zhang}, \citenamefont {Chakraborty},\ and\ \citenamefont {Behringer}}]{Bi2011}%
  \BibitemOpen
  \bibfield  {author} {\bibinfo {author} {\bibfnamefont {D.}~\bibnamefont {Bi}}, \bibinfo {author} {\bibfnamefont {J.}~\bibnamefont {Zhang}}, \bibinfo {author} {\bibfnamefont {B.}~\bibnamefont {Chakraborty}},\ and\ \bibinfo {author} {\bibfnamefont {R.~P.}\ \bibnamefont {Behringer}},\ }\bibfield  {title} {\bibinfo {title} {Jamming by shear},\ }\href {https://doi.org/10.1038/nature10667} {\bibfield  {journal} {\bibinfo  {journal} {Nature}\ }\textbf {\bibinfo {volume} {480}},\ \bibinfo {pages} {7377} (\bibinfo {year} {2011})}\BibitemShut {NoStop}%
\bibitem [{\citenamefont {Majmudar}\ \emph {et~al.}(2007)\citenamefont {Majmudar}, \citenamefont {Sperl}, \citenamefont {Luding},\ and\ \citenamefont {Behringer}}]{majmudarJammingTransitionGranular2007}%
  \BibitemOpen
  \bibfield  {author} {\bibinfo {author} {\bibfnamefont {T.~S.}\ \bibnamefont {Majmudar}}, \bibinfo {author} {\bibfnamefont {M.}~\bibnamefont {Sperl}}, \bibinfo {author} {\bibfnamefont {S.}~\bibnamefont {Luding}},\ and\ \bibinfo {author} {\bibfnamefont {R.~P.}\ \bibnamefont {Behringer}},\ }\bibfield  {title} {\bibinfo {title} {Jamming {{Transition}} in {{Granular Systems}}},\ }\href {https://doi.org/10.1103/PhysRevLett.98.058001} {\bibfield  {journal} {\bibinfo  {journal} {Phys. Rev. Lett.}\ }\textbf {\bibinfo {volume} {98}},\ \bibinfo {pages} {058001} (\bibinfo {year} {2007})}\BibitemShut {NoStop}%
\bibitem [{\citenamefont {Brown}\ \emph {et~al.}(2010)\citenamefont {Brown}, \citenamefont {Rodenberg}, \citenamefont {Amend}, \citenamefont {Mozeika}, \citenamefont {Steltz}, \citenamefont {Zakin}, \citenamefont {Lipson},\ and\ \citenamefont {Jaeger}}]{Brown2010}%
  \BibitemOpen
  \bibfield  {author} {\bibinfo {author} {\bibfnamefont {E.}~\bibnamefont {Brown}}, \bibinfo {author} {\bibfnamefont {N.}~\bibnamefont {Rodenberg}}, \bibinfo {author} {\bibfnamefont {J.}~\bibnamefont {Amend}}, \bibinfo {author} {\bibfnamefont {A.}~\bibnamefont {Mozeika}}, \bibinfo {author} {\bibfnamefont {E.}~\bibnamefont {Steltz}}, \bibinfo {author} {\bibfnamefont {M.~R.}\ \bibnamefont {Zakin}}, \bibinfo {author} {\bibfnamefont {H.}~\bibnamefont {Lipson}},\ and\ \bibinfo {author} {\bibfnamefont {H.~M.}\ \bibnamefont {Jaeger}},\ }\bibfield  {title} {\bibinfo {title} {Universal robotic gripper based on the jamming of granular material},\ }\href {https://doi.org/10.1073/pnas.1003250107} {\bibfield  {journal} {\bibinfo  {journal} {PNAS}\ }\textbf {\bibinfo {volume} {107}},\ \bibinfo {pages} {18809} (\bibinfo {year} {2010})}\BibitemShut {NoStop}%
\bibitem [{\citenamefont {G{\"o}tz}\ \emph {et~al.}(2022)\citenamefont {G{\"o}tz}, \citenamefont {Santarossa}, \citenamefont {Sack}, \citenamefont {P{\"o}schel},\ and\ \citenamefont {M{\"u}ller}}]{Goetz2022}%
  \BibitemOpen
  \bibfield  {author} {\bibinfo {author} {\bibfnamefont {H.}~\bibnamefont {G{\"o}tz}}, \bibinfo {author} {\bibfnamefont {A.}~\bibnamefont {Santarossa}}, \bibinfo {author} {\bibfnamefont {A.}~\bibnamefont {Sack}}, \bibinfo {author} {\bibfnamefont {T.}~\bibnamefont {P{\"o}schel}},\ and\ \bibinfo {author} {\bibfnamefont {P.}~\bibnamefont {M{\"u}ller}},\ }\bibfield  {title} {\bibinfo {title} {Soft particles reinforce robotic grippers: robotic grippers based on granular jamming of soft particles},\ }\bibfield  {journal} {\bibinfo  {journal} {Granular Matter}\ }\textbf {\bibinfo {volume} {24}},\ \href {https://doi.org/10.1007/s10035-021-01193-4} {10.1007/s10035-021-01193-4} (\bibinfo {year} {2022})\BibitemShut {NoStop}%
\bibitem [{\citenamefont {Jiang}\ \emph {et~al.}(2019)\citenamefont {Jiang}, \citenamefont {Yang}, \citenamefont {Chen},\ and\ \citenamefont {Chen}}]{jiangVariableStiffnessGripper2019}%
  \BibitemOpen
  \bibfield  {author} {\bibinfo {author} {\bibfnamefont {P.}~\bibnamefont {Jiang}}, \bibinfo {author} {\bibfnamefont {Y.}~\bibnamefont {Yang}}, \bibinfo {author} {\bibfnamefont {M.~Z.~Q.}\ \bibnamefont {Chen}},\ and\ \bibinfo {author} {\bibfnamefont {Y.}~\bibnamefont {Chen}},\ }\bibfield  {title} {\bibinfo {title} {A variable stiffness gripper based on differential drive particle jamming},\ }\href {https://doi.org/10.1088/1748-3190/ab04d1} {\bibfield  {journal} {\bibinfo  {journal} {Bioinspir. Biomim.}\ }\textbf {\bibinfo {volume} {14}},\ \bibinfo {pages} {036009} (\bibinfo {year} {2019})}\BibitemShut {NoStop}%
\bibitem [{\citenamefont {Santarossa}\ \emph {et~al.}(2023)\citenamefont {Santarossa}, \citenamefont {D'Angelo}, \citenamefont {Sack},\ and\ \citenamefont {P{\"o}schel}}]{Santarossa2023}%
  \BibitemOpen
  \bibfield  {author} {\bibinfo {author} {\bibfnamefont {A.}~\bibnamefont {Santarossa}}, \bibinfo {author} {\bibfnamefont {O.}~\bibnamefont {D'Angelo}}, \bibinfo {author} {\bibfnamefont {A.}~\bibnamefont {Sack}},\ and\ \bibinfo {author} {\bibfnamefont {T.}~\bibnamefont {P{\"o}schel}},\ }\bibfield  {title} {\bibinfo {title} {Effect of particle size on the suction mechanism in granular grippers},\ }\href {https://doi.org/10.1007/s10035-022-01306-7} {\bibfield  {journal} {\bibinfo  {journal} {Granular Matter}\ }\textbf {\bibinfo {volume} {25}},\ \bibinfo {pages} {16} (\bibinfo {year} {2023})}\BibitemShut {NoStop}%
\bibitem [{\citenamefont {Huijben}\ \emph {et~al.}(2011)\citenamefont {Huijben}, \citenamefont {Van~Herwijnen},\ and\ \citenamefont {Nijsse}}]{Huijben2011}%
  \BibitemOpen
  \bibfield  {author} {\bibinfo {author} {\bibfnamefont {F.}~\bibnamefont {Huijben}}, \bibinfo {author} {\bibfnamefont {F.}~\bibnamefont {Van~Herwijnen}},\ and\ \bibinfo {author} {\bibfnamefont {R.}~\bibnamefont {Nijsse}},\ }\bibfield  {title} {\bibinfo {title} {Concrete shell structures revisited: Introducing a new `low-tech' construction method using vacuumatics formwork},\ }in\ \href@noop {} {\emph {\bibinfo {booktitle} {Structural Membranes 2011: Proceedings of the 5\textsuperscript{th} International Conference on Textile Composites and Inflatable Structures}}}\ (\bibinfo {address} {Barcelona, Spain},\ \bibinfo {year} {2011})\BibitemShut {NoStop}%
\bibitem [{\citenamefont {Huijben}(2014)}]{Huijben2014}%
  \BibitemOpen
  \bibfield  {author} {\bibinfo {author} {\bibfnamefont {F.}~\bibnamefont {Huijben}},\ }\emph {\bibinfo {title} {Vacuumatics: 3D formwork systems: Investigations of the structural and morphological nature of vacuumatic structures so as to be used as semi-rigid formwork systems for producing 'free forms' and customised surface textures in concrete for architectural applications}},\ \href {https://doi.org/10.6100/IR775509} {Ph.D. thesis},\ \bibinfo  {school} {Technische Universiteit Eindhoven, The Netherland} (\bibinfo {year} {2014})\BibitemShut {NoStop}%
\bibitem [{\citenamefont {Brigido}\ \emph {et~al.}(2022)\citenamefont {Brigido}, \citenamefont {Burrow}, \citenamefont {Woods}, \citenamefont {Bartkowski},\ and\ \citenamefont {Zalewski}}]{Brigido2022}%
  \BibitemOpen
  \bibfield  {author} {\bibinfo {author} {\bibfnamefont {J.~D.}\ \bibnamefont {Brigido}}, \bibinfo {author} {\bibfnamefont {S.~G.}\ \bibnamefont {Burrow}}, \bibinfo {author} {\bibfnamefont {B.~K.}\ \bibnamefont {Woods}}, \bibinfo {author} {\bibfnamefont {P.}~\bibnamefont {Bartkowski}},\ and\ \bibinfo {author} {\bibfnamefont {R.}~\bibnamefont {Zalewski}} (\bibinfo {collaboration} {Bristol-Warsaw Collaboration}),\ }\bibfield  {title} {\bibinfo {title} {Flexural models for vacuum-packed particles as a variable-stiffness mechanism in smart structures},\ }\href {https://doi.org/10.1103/PhysRevApplied.17.044018} {\bibfield  {journal} {\bibinfo  {journal} {Phys. Rev. Applied}\ }\textbf {\bibinfo {volume} {17}},\ \bibinfo {pages} {044018} (\bibinfo {year} {2022})}\BibitemShut {NoStop}%
\bibitem [{\citenamefont {Cianchetti}\ \emph {et~al.}(2013)\citenamefont {Cianchetti}, \citenamefont {Ranzani}, \citenamefont {Gerboni}, \citenamefont {De~Falco}, \citenamefont {Laschi},\ and\ \citenamefont {Menciassi}}]{cianchettiSTIFFFLOPSurgicalManipulator2013}%
  \BibitemOpen
  \bibfield  {author} {\bibinfo {author} {\bibfnamefont {M.}~\bibnamefont {Cianchetti}}, \bibinfo {author} {\bibfnamefont {T.}~\bibnamefont {Ranzani}}, \bibinfo {author} {\bibfnamefont {G.}~\bibnamefont {Gerboni}}, \bibinfo {author} {\bibfnamefont {I.}~\bibnamefont {De~Falco}}, \bibinfo {author} {\bibfnamefont {C.}~\bibnamefont {Laschi}},\ and\ \bibinfo {author} {\bibfnamefont {A.}~\bibnamefont {Menciassi}},\ }\bibfield  {title} {\bibinfo {title} {{{STIFF-FLOP}} surgical manipulator: {{Mechanical}} design and experimental characterization of the single module},\ }in\ \href {https://doi.org/10.1109/IROS.2013.6696866} {\emph {\bibinfo {booktitle} {2013 {{IEEERSJ Int}}. {{Conf}}. {{Intell}}. {{Robots Syst}}.}}}\ (\bibinfo {year} {2013})\ pp.\ \bibinfo {pages} {3576--3581}\BibitemShut {NoStop}%
\bibitem [{\citenamefont {Brigido-González}\ \emph {et~al.}(2019)\citenamefont {Brigido-González}, \citenamefont {Burrow},\ and\ \citenamefont {Woods}}]{brigido-gonzalezSwitchableStiffnessMorphing2019}%
  \BibitemOpen
  \bibfield  {author} {\bibinfo {author} {\bibfnamefont {J.~D.}\ \bibnamefont {Brigido-González}}, \bibinfo {author} {\bibfnamefont {S.~G.}\ \bibnamefont {Burrow}},\ and\ \bibinfo {author} {\bibfnamefont {B.~K.}\ \bibnamefont {Woods}},\ }\bibfield  {title} {\bibinfo {title} {Switchable stiffness morphing aerostructures based on granular jamming},\ }\href {https://doi.org/10.1177/1045389X19862372} {\bibfield  {journal} {\bibinfo  {journal} {Journal of Intelligent Material Systems and Structures}\ }\textbf {\bibinfo {volume} {30}},\ \bibinfo {pages} {2581} (\bibinfo {year} {2019})}\BibitemShut {NoStop}%
\bibitem [{AST(2020)}]{ASTM2020_bending}%
  \BibitemOpen
  \href {https://doi.org/10.1520/D6272-17E01} {\emph {\bibinfo {title} {Standard Test Method for Flexural Properties of Unreinforced and Reinforced Plastics and Electrical Insulating Materials by Four-Point Bending}}},\ \bibinfo {type} {Standard}\ \bibinfo {number} {ASTM D6272-17e1}\ (\bibinfo  {institution} {ASTM International},\ \bibinfo {year} {2020})\BibitemShut {NoStop}%
\bibitem [{DIN(2011)}]{DINENISO2011_bending}%
  \BibitemOpen
  \href {https://doi.org/10.31030/3344826} {\emph {\bibinfo {title} {Fibre-reinforced plastic composites -- Determination of flexural properties}}},\ \bibinfo {type} {Standard}\ \bibinfo {number} {DIN EN ISO 14125}\ (\bibinfo  {institution} {Deutsches Institut für Normung e.V.},\ \bibinfo {year} {2011})\BibitemShut {NoStop}%
\bibitem [{\citenamefont {Liu}\ and\ \citenamefont {Nagel}(1993)}]{liuSoundGranularMaterial1993}%
  \BibitemOpen
  \bibfield  {author} {\bibinfo {author} {\bibfnamefont {C.-h.}\ \bibnamefont {Liu}}\ and\ \bibinfo {author} {\bibfnamefont {S.~R.}\ \bibnamefont {Nagel}},\ }\bibfield  {title} {\bibinfo {title} {Sound in a granular material: Disorder and nonlinearity},\ }\href {https://doi.org/10.1103/PhysRevB.48.15646} {\bibfield  {journal} {\bibinfo  {journal} {Physical Review B}\ }\textbf {\bibinfo {volume} {48}},\ \bibinfo {pages} {15646} (\bibinfo {year} {1993})}\BibitemShut {NoStop}%
\bibitem [{\citenamefont {Tell}\ \emph {et~al.}(2020)\citenamefont {Tell}, \citenamefont {Drei{\ss}igacker}, \citenamefont {Tchapnda}, \citenamefont {Yu},\ and\ \citenamefont {Sperl}}]{tellAcousticWavesGranular2020}%
  \BibitemOpen
  \bibfield  {author} {\bibinfo {author} {\bibfnamefont {K.}~\bibnamefont {Tell}}, \bibinfo {author} {\bibfnamefont {C.}~\bibnamefont {Drei{\ss}igacker}}, \bibinfo {author} {\bibfnamefont {A.~C.}\ \bibnamefont {Tchapnda}}, \bibinfo {author} {\bibfnamefont {P.}~\bibnamefont {Yu}},\ and\ \bibinfo {author} {\bibfnamefont {M.}~\bibnamefont {Sperl}},\ }\bibfield  {title} {\bibinfo {title} {Acoustic waves in granular packings at low confinement pressure},\ }\href {https://doi.org/10.1063/1.5122848} {\bibfield  {journal} {\bibinfo  {journal} {Review of Scientific Instruments}\ }\textbf {\bibinfo {volume} {91}},\ \bibinfo {pages} {033906} (\bibinfo {year} {2020})}\BibitemShut {NoStop}%
\bibitem [{\citenamefont {P{\"o}schel}\ and\ \citenamefont {Schwager}(2005)}]{poschelComputationalGranularDynamics2005}%
  \BibitemOpen
  \bibfield  {author} {\bibinfo {author} {\bibfnamefont {T.}~\bibnamefont {P{\"o}schel}}\ and\ \bibinfo {author} {\bibfnamefont {T.}~\bibnamefont {Schwager}},\ }\href {https://doi.org/10.1007/3-540-27720-X} {\emph {\bibinfo {title} {Computational Granular Dynamics: Models and Algorithms}}}\ (\bibinfo  {publisher} {Springer},\ \bibinfo {address} {Berlin},\ \bibinfo {year} {2005})\BibitemShut {NoStop}%
\bibitem [{\citenamefont {Matuttis}\ and\ \citenamefont {Chen}(2014)}]{matuttisUnderstandingDiscreteElement2014}%
  \BibitemOpen
  \bibfield  {author} {\bibinfo {author} {\bibfnamefont {H.-G.}\ \bibnamefont {Matuttis}}\ and\ \bibinfo {author} {\bibfnamefont {J.}~\bibnamefont {Chen}},\ }\href {https://doi.org/10.1002/9781118567210} {\emph {\bibinfo {title} {Understanding the Discrete Element Method: Simulation of Non-Spherical Particles for Granular and Multi-body Systems}}}\ (\bibinfo  {publisher} {Wiley},\ \bibinfo {address} {Singapore},\ \bibinfo {year} {2014})\BibitemShut {NoStop}%
\bibitem [{\citenamefont {Luding}(2008)}]{ludingIntroductionDiscreteElement2008}%
  \BibitemOpen
  \bibfield  {author} {\bibinfo {author} {\bibfnamefont {S.}~\bibnamefont {Luding}},\ }\bibfield  {title} {\bibinfo {title} {Introduction to discrete element methods: {{Basic}} of contact force models and how to perform the micro-macro transition to continuum theory},\ }\href {https://doi.org/10.1080/19648189.2008.9693050} {\bibfield  {journal} {\bibinfo  {journal} {European Journal of Environmental and Civil Engineering}\ }\textbf {\bibinfo {volume} {12}},\ \bibinfo {pages} {785} (\bibinfo {year} {2008})}\BibitemShut {NoStop}%
\bibitem [{\citenamefont {Thornton}\ \emph {et~al.}(2023)\citenamefont {Thornton}, \citenamefont {Plath}, \citenamefont {Ostanin}, \citenamefont {G{\"o}tz}, \citenamefont {Bisschop}, \citenamefont {Hassan}, \citenamefont {Roeplal}, \citenamefont {Wang}, \citenamefont {Pourandi},\ and\ \citenamefont {Weinhart}}]{thorntonRecentAdvancesMercuryDPM2023}%
  \BibitemOpen
  \bibfield  {author} {\bibinfo {author} {\bibfnamefont {A.~R.}\ \bibnamefont {Thornton}}, \bibinfo {author} {\bibfnamefont {T.}~\bibnamefont {Plath}}, \bibinfo {author} {\bibfnamefont {I.}~\bibnamefont {Ostanin}}, \bibinfo {author} {\bibfnamefont {H.}~\bibnamefont {G{\"o}tz}}, \bibinfo {author} {\bibfnamefont {J.-W.}\ \bibnamefont {Bisschop}}, \bibinfo {author} {\bibfnamefont {M.}~\bibnamefont {Hassan}}, \bibinfo {author} {\bibfnamefont {R.}~\bibnamefont {Roeplal}}, \bibinfo {author} {\bibfnamefont {X.}~\bibnamefont {Wang}}, \bibinfo {author} {\bibfnamefont {S.}~\bibnamefont {Pourandi}},\ and\ \bibinfo {author} {\bibfnamefont {T.}~\bibnamefont {Weinhart}},\ }\bibfield  {title} {\bibinfo {title} {Recent {{Advances}} in {{MercuryDPM}}},\ }\href {https://doi.org/10.1007/s11786-023-00562-x} {\bibfield  {journal} {\bibinfo  {journal} {Math.Comput.Sci.}\ }\textbf {\bibinfo {volume} {17}},\ \bibinfo {pages} {13} (\bibinfo {year} {2023})}\BibitemShut {NoStop}%
\bibitem [{\citenamefont {G{\"o}tz}\ and\ \citenamefont {P{\"o}schel}(2023{\natexlab{b}})}]{goetzDEMSimulationThinElastic2022}%
  \BibitemOpen
  \bibfield  {author} {\bibinfo {author} {\bibfnamefont {H.}~\bibnamefont {G{\"o}tz}}\ and\ \bibinfo {author} {\bibfnamefont {T.}~\bibnamefont {P{\"o}schel}},\ }\bibfield  {title} {\bibinfo {title} {{{DEM-simulation}} of thin elastic membranes interacting with a granulate},\ }\href {https://doi.org/10.1007/s10035-023-01344-9} {\bibfield  {journal} {\bibinfo  {journal} {Granular Matter}\ }\textbf {\bibinfo {volume} {25}},\ \bibinfo {pages} {61} (\bibinfo {year} {2023}{\natexlab{b}})}\BibitemShut {NoStop}%
\bibitem [{\citenamefont {Lubachevsky}\ and\ \citenamefont {Stillinger}(1990)}]{lubachevskyGeometricPropertiesRandom1990}%
  \BibitemOpen
  \bibfield  {author} {\bibinfo {author} {\bibfnamefont {B.~D.}\ \bibnamefont {Lubachevsky}}\ and\ \bibinfo {author} {\bibfnamefont {F.~H.}\ \bibnamefont {Stillinger}},\ }\bibfield  {title} {\bibinfo {title} {Geometric properties of random disk packings},\ }\href {https://doi.org/10.1007/BF01025983} {\bibfield  {journal} {\bibinfo  {journal} {J Stat Phys}\ }\textbf {\bibinfo {volume} {60}},\ \bibinfo {pages} {561} (\bibinfo {year} {1990})}\BibitemShut {NoStop}%
\bibitem [{\citenamefont {Song}\ \emph {et~al.}(2008)\citenamefont {Song}, \citenamefont {Wang},\ and\ \citenamefont {Makse}}]{Song2008}%
  \BibitemOpen
  \bibfield  {author} {\bibinfo {author} {\bibfnamefont {C.}~\bibnamefont {Song}}, \bibinfo {author} {\bibfnamefont {P.}~\bibnamefont {Wang}},\ and\ \bibinfo {author} {\bibfnamefont {H.~A.}\ \bibnamefont {Makse}},\ }\bibfield  {title} {\bibinfo {title} {A phase diagram for jammed matter},\ }\href {https://doi.org/10.1038/nature06981} {\bibfield  {journal} {\bibinfo  {journal} {Nature}\ }\textbf {\bibinfo {volume} {453}},\ \bibinfo {pages} {7195} (\bibinfo {year} {2008})}\BibitemShut {NoStop}%
\bibitem [{\citenamefont {Jin}\ and\ \citenamefont {Makse}(2010)}]{jinFirstorderPhaseTransition2010}%
  \BibitemOpen
  \bibfield  {author} {\bibinfo {author} {\bibfnamefont {Y.}~\bibnamefont {Jin}}\ and\ \bibinfo {author} {\bibfnamefont {H.~A.}\ \bibnamefont {Makse}},\ }\bibfield  {title} {\bibinfo {title} {A first-order phase transition defines the random close packing of hard spheres},\ }\href {https://doi.org/10.1016/j.physa.2010.08.010} {\bibfield  {journal} {\bibinfo  {journal} {Physica A: Statistical Mechanics and its Applications}\ }\textbf {\bibinfo {volume} {389}},\ \bibinfo {pages} {5362} (\bibinfo {year} {2010})}\BibitemShut {NoStop}%
\bibitem [{\citenamefont {Truesdell}(1960)}]{truesdellRationalMechanicsFlexible1960}%
  \BibitemOpen
  \bibfield  {author} {\bibinfo {author} {\bibfnamefont {C.}~\bibnamefont {Truesdell}},\ }\href@noop {} {\emph {\bibinfo {title} {The Rational Mechanics of Flexible or Elastic Bodies 1638-1788. {{Introduction}} to {{Leonhardi}} Euleri Opera Omnia {{Vol}}. {{X}} et {{XI}} Series Secundae}}}\ (\bibinfo  {publisher} {{Birkh\"auser Basel}},\ \bibinfo {year} {1960})\BibitemShut {NoStop}%
\bibitem [{\citenamefont {Falk}\ and\ \citenamefont {Langer}(1998)}]{falkDynamicsViscoplasticDeformation1998}%
  \BibitemOpen
  \bibfield  {author} {\bibinfo {author} {\bibfnamefont {M.~L.}\ \bibnamefont {Falk}}\ and\ \bibinfo {author} {\bibfnamefont {J.~S.}\ \bibnamefont {Langer}},\ }\bibfield  {title} {\bibinfo {title} {Dynamics of viscoplastic deformation in amorphous solids},\ }\href {https://doi.org/10.1103/PhysRevE.57.7192} {\bibfield  {journal} {\bibinfo  {journal} {Phys. Rev. E}\ }\textbf {\bibinfo {volume} {57}},\ \bibinfo {pages} {7192} (\bibinfo {year} {1998})}\BibitemShut {NoStop}%
\bibitem [{\citenamefont {Perrin}\ \emph {et~al.}(2019)\citenamefont {Perrin}, \citenamefont {Clavaud}, \citenamefont {Wyart}, \citenamefont {Metzger},\ and\ \citenamefont {Forterre}}]{Perrin2019}%
  \BibitemOpen
  \bibfield  {author} {\bibinfo {author} {\bibfnamefont {H.}~\bibnamefont {Perrin}}, \bibinfo {author} {\bibfnamefont {C.}~\bibnamefont {Clavaud}}, \bibinfo {author} {\bibfnamefont {M.}~\bibnamefont {Wyart}}, \bibinfo {author} {\bibfnamefont {B.}~\bibnamefont {Metzger}},\ and\ \bibinfo {author} {\bibfnamefont {Y.}~\bibnamefont {Forterre}},\ }\bibfield  {title} {\bibinfo {title} {Interparticle friction leads to nonmonotonic flow curves and hysteresis in viscous suspensions},\ }\href {https://doi.org/10.1103/PhysRevX.9.031027} {\bibfield  {journal} {\bibinfo  {journal} {Physical Review X}\ }\textbf {\bibinfo {volume} {9}},\ \bibinfo {pages} {031027} (\bibinfo {year} {2019})}\BibitemShut {NoStop}%
\bibitem [{\citenamefont {Rudge}\ \emph {et~al.}(2020)\citenamefont {Rudge}, \citenamefont {Scholten},\ and\ \citenamefont {Dijksman}}]{Rudge2020}%
  \BibitemOpen
  \bibfield  {author} {\bibinfo {author} {\bibfnamefont {R.~E.}\ \bibnamefont {Rudge}}, \bibinfo {author} {\bibfnamefont {E.}~\bibnamefont {Scholten}},\ and\ \bibinfo {author} {\bibfnamefont {J.~A.}\ \bibnamefont {Dijksman}},\ }\bibfield  {title} {\bibinfo {title} {Natural and induced surface roughness determine frictional regimes in hydrogel pairs},\ }\href {https://doi.org/10.1016/j.triboint.2019.105903} {\bibfield  {journal} {\bibinfo  {journal} {Tribology International}\ }\textbf {\bibinfo {volume} {141}},\ \bibinfo {pages} {105903} (\bibinfo {year} {2020})}\BibitemShut {NoStop}%
\bibitem [{\citenamefont {Sch{\"a}fer}\ \emph {et~al.}(1996)\citenamefont {Sch{\"a}fer}, \citenamefont {Dippel},\ and\ \citenamefont {Wolf}}]{Shafer1996}%
  \BibitemOpen
  \bibfield  {author} {\bibinfo {author} {\bibfnamefont {J.}~\bibnamefont {Sch{\"a}fer}}, \bibinfo {author} {\bibfnamefont {S.}~\bibnamefont {Dippel}},\ and\ \bibinfo {author} {\bibfnamefont {D.~E.}\ \bibnamefont {Wolf}},\ }\bibfield  {title} {\bibinfo {title} {Force {{Schemes}} in {{Simulations}} of {{Granular Materials}}},\ }\href {https://doi.org/10.1051/jp1:1996129} {\bibfield  {journal} {\bibinfo  {journal} {J. Phys. I France}\ }\textbf {\bibinfo {volume} {6}},\ \bibinfo {pages} {5} (\bibinfo {year} {1996})}\BibitemShut {NoStop}%
\bibitem [{\citenamefont {{Kruggel-Emden}}\ \emph {et~al.}(2007)\citenamefont {{Kruggel-Emden}}, \citenamefont {Simsek}, \citenamefont {Rickelt}, \citenamefont {Wirtz},\ and\ \citenamefont {Scherer}}]{kruggel-emdenReviewExtensionNormal2007}%
  \BibitemOpen
  \bibfield  {author} {\bibinfo {author} {\bibfnamefont {H.}~\bibnamefont {{Kruggel-Emden}}}, \bibinfo {author} {\bibfnamefont {E.}~\bibnamefont {Simsek}}, \bibinfo {author} {\bibfnamefont {S.}~\bibnamefont {Rickelt}}, \bibinfo {author} {\bibfnamefont {S.}~\bibnamefont {Wirtz}},\ and\ \bibinfo {author} {\bibfnamefont {V.}~\bibnamefont {Scherer}},\ }\bibfield  {title} {\bibinfo {title} {Review and extension of normal force models for the {{Discrete Element Method}}},\ }\href {https://doi.org/10.1016/j.powtec.2006.10.004} {\bibfield  {journal} {\bibinfo  {journal} {Powder Technology}\ }\textbf {\bibinfo {volume} {171}},\ \bibinfo {pages} {157} (\bibinfo {year} {2007})}\BibitemShut {NoStop}%
\bibitem [{\citenamefont {Brilliantov}\ \emph {et~al.}(1996)\citenamefont {Brilliantov}, \citenamefont {Spahn}, \citenamefont {Hertzsch},\ and\ \citenamefont {P{\"o}schel}}]{brilliantovModelCollisionsGranular1996}%
  \BibitemOpen
  \bibfield  {author} {\bibinfo {author} {\bibfnamefont {N.~V.}\ \bibnamefont {Brilliantov}}, \bibinfo {author} {\bibfnamefont {F.}~\bibnamefont {Spahn}}, \bibinfo {author} {\bibfnamefont {J.-M.}\ \bibnamefont {Hertzsch}},\ and\ \bibinfo {author} {\bibfnamefont {T.}~\bibnamefont {P{\"o}schel}},\ }\bibfield  {title} {\bibinfo {title} {Model for collisions in granular gases},\ }\href {https://doi.org/10.1103/PhysRevE.53.5382} {\bibfield  {journal} {\bibinfo  {journal} {Phys. Rev. E}\ }\textbf {\bibinfo {volume} {53}},\ \bibinfo {pages} {5382} (\bibinfo {year} {1996})}\BibitemShut {NoStop}%
\bibitem [{\citenamefont {Kuwabara}\ and\ \citenamefont {Kono}(1987)}]{kuwabaraRestitutionCoefficientCollision1987}%
  \BibitemOpen
  \bibfield  {author} {\bibinfo {author} {\bibfnamefont {G.}~\bibnamefont {Kuwabara}}\ and\ \bibinfo {author} {\bibfnamefont {K.}~\bibnamefont {Kono}},\ }\bibfield  {title} {\bibinfo {title} {Restitution {{Coefficient}} in a {{Collision}} between {{Two Spheres}}},\ }\href {https://doi.org/10.1143/JJAP.26.1230} {\bibfield  {journal} {\bibinfo  {journal} {Japanese Journal of Applied Physics}\ }\textbf {\bibinfo {volume} {26}},\ \bibinfo {pages} {1230} (\bibinfo {year} {1987})}\BibitemShut {NoStop}%
\bibitem [{\citenamefont {Müller}\ and\ \citenamefont {Pöschel}(2011)}]{mullerCollisionViscoelasticSpheres2011}%
  \BibitemOpen
  \bibfield  {author} {\bibinfo {author} {\bibfnamefont {P.}~\bibnamefont {Müller}}\ and\ \bibinfo {author} {\bibfnamefont {T.}~\bibnamefont {Pöschel}},\ }\bibfield  {title} {\bibinfo {title} {Collision of viscoelastic spheres: {{Compact}} expressions for the coefficient of normal restitution},\ }\href {https://doi.org/10.1103/PhysRevE.84.021302} {\bibfield  {journal} {\bibinfo  {journal} {Phys. Rev. E}\ }\textbf {\bibinfo {volume} {84}},\ \bibinfo {pages} {021302} (\bibinfo {year} {2011})}\BibitemShut {NoStop}%
\bibitem [{\citenamefont {{Kruggel-Emden}}\ \emph {et~al.}(2008)\citenamefont {{Kruggel-Emden}}, \citenamefont {Wirtz},\ and\ \citenamefont {Scherer}}]{kruggel-emdenStudyTangentialForce2008}%
  \BibitemOpen
  \bibfield  {author} {\bibinfo {author} {\bibfnamefont {H.}~\bibnamefont {{Kruggel-Emden}}}, \bibinfo {author} {\bibfnamefont {S.}~\bibnamefont {Wirtz}},\ and\ \bibinfo {author} {\bibfnamefont {V.}~\bibnamefont {Scherer}},\ }\bibfield  {title} {\bibinfo {title} {A study on tangential force laws applicable to the discrete element method ({{DEM}}) for materials with viscoelastic or plastic behavior},\ }\href {https://doi.org/10.1016/j.ces.2007.11.025} {\bibfield  {journal} {\bibinfo  {journal} {Chem. Eng. Sci.}\ }\textbf {\bibinfo {volume} {63}},\ \bibinfo {pages} {1523} (\bibinfo {year} {2008})}\BibitemShut {NoStop}%
\bibitem [{\citenamefont {Kot}\ and\ \citenamefont {Nagahashi}(2017)}]{kotMassSpringModels2017}%
  \BibitemOpen
  \bibfield  {author} {\bibinfo {author} {\bibfnamefont {M.}~\bibnamefont {Kot}}\ and\ \bibinfo {author} {\bibfnamefont {H.}~\bibnamefont {Nagahashi}},\ }\bibfield  {title} {\bibinfo {title} {Mass spring models with adjustable {{Poisson}}'s ratio},\ }\href {https://doi.org/10.1007/s00371-015-1194-8} {\bibfield  {journal} {\bibinfo  {journal} {Vis Comput}\ }\textbf {\bibinfo {volume} {33}},\ \bibinfo {pages} {283} (\bibinfo {year} {2017})}\BibitemShut {NoStop}%
\bibitem [{\citenamefont {Goldhirsch}(2010)}]{goldhirschStressStressAsymmetry2010}%
  \BibitemOpen
  \bibfield  {author} {\bibinfo {author} {\bibfnamefont {I.}~\bibnamefont {Goldhirsch}},\ }\bibfield  {title} {\bibinfo {title} {Stress, stress asymmetry and couple stress: From discrete particles to continuous fields},\ }\href {https://doi.org/10.1007/s10035-010-0181-z} {\bibfield  {journal} {\bibinfo  {journal} {Granular Matter}\ }\textbf {\bibinfo {volume} {12}},\ \bibinfo {pages} {239} (\bibinfo {year} {2010})}\BibitemShut {NoStop}%
\bibitem [{\citenamefont {Weinhart}\ \emph {et~al.}(2016)\citenamefont {Weinhart}, \citenamefont {Labra}, \citenamefont {Luding},\ and\ \citenamefont {Ooi}}]{weinhartInfluenceCoarsegrainingParameters2016}%
  \BibitemOpen
  \bibfield  {author} {\bibinfo {author} {\bibfnamefont {T.}~\bibnamefont {Weinhart}}, \bibinfo {author} {\bibfnamefont {C.}~\bibnamefont {Labra}}, \bibinfo {author} {\bibfnamefont {S.}~\bibnamefont {Luding}},\ and\ \bibinfo {author} {\bibfnamefont {J.~Y.}\ \bibnamefont {Ooi}},\ }\bibfield  {title} {\bibinfo {title} {Influence of coarse-graining parameters on the analysis of {{DEM}} simulations of silo flow},\ }\href {https://doi.org/10.1016/j.powtec.2015.11.052} {\bibfield  {journal} {\bibinfo  {journal} {Powder Technology}\ }\bibinfo {series} {Particle {{Modelling}} with the {{Discrete Element Method A}} Success Story of {{PARDEM}} (www.pardem.eu)},\ \textbf {\bibinfo {volume} {293}},\ \bibinfo {pages} {138} (\bibinfo {year} {2016})}\BibitemShut {NoStop}%
\end{thebibliography}%

\appendix
\section{Simulation Method}
\label{sec:methods}

\subsection{Particle-particle interaction}
\label{sec:methods:particle-particle}

For the particle-particle interaction, we assume viscoelastic spheres with dry friction  \cite{Shafer1996,kruggel-emdenReviewExtensionNormal2007}.
Two contacting particles at positions $\vec{r}_i$ and $\vec{r}_j$ \red{with radii $R_i$ and $R_j$, and masses $m_i$, $m_j$}, feel the normal force
\begin{align}
  \vec{F}^\text{n}_{ij} = \max\left(0, \frac{2E_\text{p}\sqrt{R_{ij}^\text{eff}}}{3(1-\nu_\text{p}^2)}\left(\xi_{ij}^\frac{3}{2} - \frac{3}{2}A\xi_{ij}^\frac{1}{2}\dot{\xi}_{ij}\right)\right)\frac{\vec{r}_i-\vec{r}_j}{|\vec{r}_i-\vec{r}_j|}\,.
\end{align}
Here, we use $\xi_{ij}=R_i + R_j - |\vec{r}_i-\vec{r}_j|$ and $R_{ij}^\text{eff}=R_iR_j/(R_i+R_j)$. The particle material is described by the elastic modulus, $E_\text{p}$, the Poisson ratio, $\nu_\text{p}$, and the \red{relaxation time} $A$, see \red{\cite{brilliantovModelCollisionsGranular1996,kuwabaraRestitutionCoefficientCollision1987}}. We use $A=\SI{7e-6}{\red{\second}}$, which corresponds to the coefficient of restitution  $\varepsilon\approx0.8$  for spheres of  radii $R=2.5\,\text{mm}$, with $E_\text{p}=100\,\text{MPa}$ and $\nu_\text{p}=0.254$  impacting at velocity  $2\,\text{m}/\text{s}$ \cite{mullerCollisionViscoelasticSpheres2011}.

Friction between non-sliding contacting particles is modeled by a force in tangential direction \cite{kruggel-emdenStudyTangentialForce2008},
\begin{align}
  \vec{F}_{ij}^\text{t} = -\min\left(\frac{2}{3}k^\text{t}_{ij}\xi^\text{t}_{ij},\,\mu F_n\right)\frac{\vec{\xi}^\text{t}_{ij}}{\xi^\text{t}_{ij}}.\label{eq:tangentialForce}
\end{align}
Here, assuming isotropic particles,
\begin{align}
  k^\text{t}_{ij} = 8\frac{\red{E_\text{p}}}{2(1+\nu_\text{p})(2-\nu_\text{p})}\sqrt{R_{ij}^\text{eff}\xi_{ij}},
\end{align}
is the tangential stiffness and $\mu$ the friction coefficient. $\vec{\xi}^\text{t}_{ij}$ is the vector valued tangential compression, which is set so zero on contact formation and integrated using 
\begin{align}
  \dot{\vec{\xi}}^\text{t}_{ij} = (\dot{\vec{r}}_i + \vec{\omega}_i\times\vec{b}_{ij}) - (\dot{\vec{r}}_j + \vec{\omega}_j\times\vec{b}_{ji}) - \dot{\xi}\frac{\vec{r}_i-\vec{r}_j}{|\vec{r}_i-\vec{r}_j|},
\end{align}
where $\vec{\omega}_i$,  $\vec{\omega}_j$ are the particles' angular velocities and $\vec{b}_{ij}=-(R_i-\xi_{ij}/2)\frac{\vec{r}_i-\vec{r}_j}{|\vec{r}_i-\vec{r}_j|}$ points from the center of particle $i$ towards the contact point. Throughout the contact, we ensure $\vec{\xi}^\text{t}_{ij}$ is tangential to the contact by rotating it appropriately \cite{ludingIntroductionDiscreteElement2008}. The minimum rule in Eq.~\eqref{eq:tangentialForce} ensures, that the Coulomb rule is upheld. If $|\vec{F}^\text{t}_{ij}|\leq\mu|\vec{F}^\text{n}_{ij}|$, we include sliding dissipation and modify the tangential force
\begin{align}
  \vec{F}_{ij}^\text{t} = -\red{\frac{2}{3}k^\text{t}_{ij}\vec{\xi}^\text{t}_{ij}} - \gamma\sqrt{m_{ij}^\text{eff}k_{ij}^\text{t}}\dot{\vec{\xi}}^\text{t}_{ij}.
\end{align}
\red{Here, $\gamma$ is twice the damping ratio and $m_{ij}^\text{eff}=m_im_j/(m_i+m_j)$ is the effective mass.}

\subsection{Particle-membrane}
We use a 2D mass-spring system to model the flexible membrane \cite{kotMassSpringModels2017} and define the connectivity of the membrane's vertex particles by a triangular mesh with a hexagonal unit cell. We insert triangular patches between the membrane's vertex particles. The interaction between a granular particle and the membrane is then given by the contacts of the granular particle with the triangular patches of the membrane. The velocity and the contact forces are interpolated between the patch and the associated vertex particles based on the barycentric coordinates of the contact point. This interpolation ensures a well-defined behavior for contacts sliding along the membrane. For a more detailed description of the membrane model, we refer to~\cite{goetzDEMSimulationThinElastic2022}.

The repulsive force of the contact is calculated similarly to the particle-particle force described in Appendix \ref{sec:methods:particle-particle}. 

\subsection{Membrane-mold interaction}
The interaction between the mold and the membrane is given by the contacts between the membrane's vertex particles and the mold. Again, the force of the contact is calculated similar to the particle-particle force described in Appendix \ref{sec:methods:particle-particle}.

\subsection{Pressure application}
To apply the confining pressure to the membrane, each triangular patch of area, $A^w_i$, and normal unit vector, $\hat{\vec{e}}^{\,w}_i$, is loaded with a force, $\vec{F}^{\,w}_i=A^{\,w}_i\Delta p\,\hat{\vec{e}}^{\,w}_i$, where $\hat{\vec{e}}^{\,w}_i$ is defined such that the force acts from outside the membrane to the granulate located inside.

\subsection{Membrane supports and beam deflection}
\begin{figure}[h!]
  \centering
\printfig{\includegraphics[width=\linewidth]{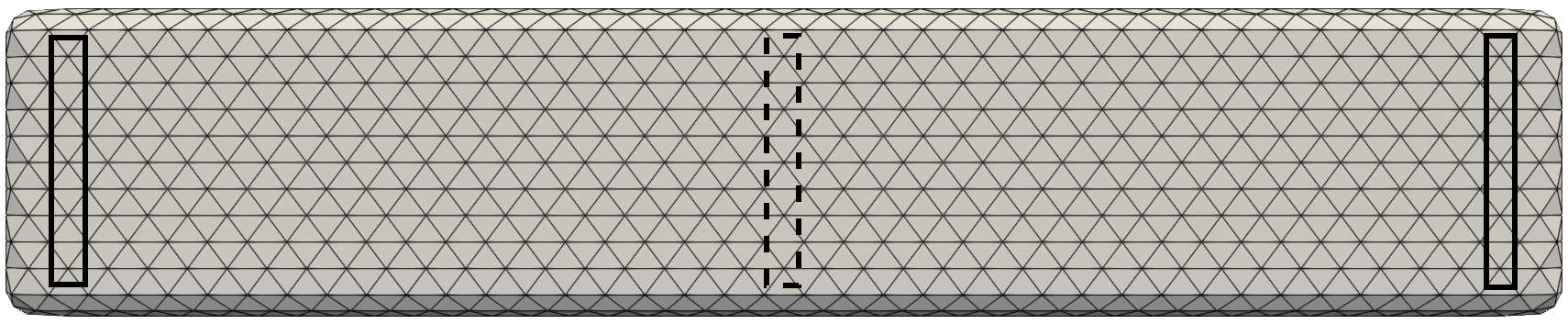}}
  \caption{Membrane enclosing the granular beam. Solid rectangles indicate the fixed vertex particles or holding points of the beam. The dashed rectangle indicates the region of vertex particles used to measure the displacement $\Delta z$.}
  \label{fig:beam_vertex_particles}
\end{figure}
We fix the translational degrees of freedom of some vertex particles close to the beam's extremities to model the simple supports used in the bending experiment.
Similarly, membrane-particles in the middle of the beam's bottom face are used to determine the beam's deflection by recording their positions. Figure~\ref{fig:beam_vertex_particles} indicates the exact location of the respective membrane-particles in our system. 

\subsection{Simulation parameters}
If not specified differently, in our simulations we use the parameters given in Tab.  \ref{tab:matProp}.
\begin{table}[ht]
  \centering
  \caption{\label{tab:matProp}
  Material parameters of the particles and the membrane used in simulations.}
  \begin{tabular}{l l l}
  \toprule
   & \textbf{particles}	& \textbf{membrane}\\
   \midrule
       radius $R$ / thickness (mm) &  $2.2$ to $2.5$ & 0.3 \\
       density $\rho$ (kg/m$^3$) & 2000 & 2000 \\
       elastic modulus \red{$E_\text{p}$} (Pa) & $5\cdot10^{6}$ to $10^{8}$ & $10^{7}$\\
       Poisson's ratio $\nu$ & 0.245 & 0.33\\
       normal viscous parameter $A$ & $7\cdot10^{-6}$ & $7\cdot10^{-6}$\\ 
    friction coefficient $\mu_\text{p}$ / $\mu_\text{m}$ & 0 to 1.2 & 1.2\\
    initial friction coefficient $\mu_\text{init}$ & 0.3 &\\
    tangential damping $\gamma$ & 0.3 & 0.3\\
    membrane damping $\gamma_m$ & & $0.15$\\
    \bottomrule
  \end{tabular}
  \end{table}
\section{Deriving stress and strain
  \label{apx:timoshenko}}

For the stiffness we consult the static Euler-Bernoulli beam theory for small deformations. The theory relates the bending moment in $y$-direction, $M_{y}$, and shear force in $z$ direction, $Q_z$, of a beam of length, $L$, with elastic modulus, $E$, to the displacement, $\Delta z$, through the following differential equations:
\begin{align}
  \begin{split}
    M_{y} &= -EI\frac{\mathrm{d}^2\Delta z}{\mathrm{d}x^2}\\
    Q_{z} &= \frac{-\mathrm{d}}{\mathrm{d}x} \left(EI\frac{\mathrm{d}^2\Delta z}{\mathrm{d}x^2}\right)
  \end{split}\label{eq:timoshenko}
\end{align}
From our setup of a simply supported four point bending beam with a load length of one third of the beam length, we can further specify
\begin{align}
  M_{y}(x) &= \frac{1}{2}\begin{cases}
    Fx & \text{for } x<L/3\\
    \frac{FL}{3} & \text{for } L/3\leq x\leq 2L/3\\
    F(L-x) & \text{for } 2L/3 < x
  \end{cases}\label{eq:moment}\\
  Q_{z}(x) &= \frac{1}{2}\begin{cases}
    F & \text{for } x<L/3\\
    0 & \text{for } L/3\leq x\leq 2L/3\\
    -F & \text{for } 2L/3 < x
  \end{cases}\label{eq:shear}.
\end{align}
We integrate Eq. (\ref{eq:timoshenko}) with Eqs. (\ref{eq:moment}) and (\ref{eq:shear}) while applying appropriate boundary conditions. At the position $x=L/2$ we obtain the following relation between displacement and applied force
\begin{align}
  \Delta z\left(\frac{L}{2}\right)=\frac{FL^3}{6}\frac{23}{108EI}.\label{eq:timoshenkoForceDeflection}
\end{align}
We calculate the stress at the position $x=L/2, z=-h/2$
\begin{align}
  \sigma = -\frac{M_yz}{I} = \frac{FLh}{6I} = \frac{FL}{h^2d}\label{eq:apx:timoshenkoStress},
\end{align}
where we used the height, $h$, and depth, $d$, of the beam. With the assumption $\sigma = E\epsilon$ at small strain, Eqs. (\ref{eq:timoshenkoForceDeflection}), and (\ref{eq:apx:timoshenkoStress}), allow us to define the strain
\begin{align}
  \epsilon = \Delta z\left(\frac{L}{2}\right)\frac{108h}{23L^2}.
\end{align}
This is approximately $\epsilon\approx 4.7\Delta z(L/2)h/L^2$, which is equivalent to the formular provided, e.g., in \cite{ASTM2020_bending}. Note, in the main text $\Delta z$ refers to the beam's central displacement, i.e., $\Delta z\equiv\Delta z(L/2)$.

\section{Coarse grained displacement}
\label{supp:displacement}
The displacement $\Delta\vec{r}_i(t)$ of particle, $i$, relative to the beam's center, $x_c$, (the position halfway between the beams extremities in $x$- and $y$-direction) is given by
\begin{align}
  \Delta\vec{r}_i(t) &= \left(\vec{r}_i(t) - \vec{r}_i(t-\Delta t)\right)\odot\frac{\vec{r}_i(t)-\vec{x}_c}{|\vec{r}_i(t)-\vec{x}_c|},
\end{align}
where $\odot$ is an element-wise multiplication. We apply coarse graining with a Gaussian kernel function with width of one particle diameter and divide by the coarse grained number density to obtain the continuum displacement $\Delta\vec{r}(\vec{r}, t)$ for any position $\vec{r}$ within the beam \cite{goldhirschStressStressAsymmetry2010,weinhartInfluenceCoarsegrainingParameters2016}.

\section{Autocorrelation function}
\label{supp:autoCorr}

We use auto correlation functions to study the spatial and temporal similarity of $D_\text{min}$ values within the beam. We calculate the spatial autocorrelation with
\begin{align}
  C_i(r) &= \frac{1}{M} \sum_j \frac{D_{\text{min},i}D_{\text{min},j}}{d_\text{mean}},\quad |\vec{r}_i-\vec{r}_j|=r\pm\Delta r,\\
  C(r) &= \frac{1}{N}\sum_i C_i(r),
\end{align}
where $C_i(r)$ is the spatial auto-correlation function for particle $i$. It takes all $M$ particles $j$ at a distance of $r\pm\Delta r$ into account. $C(r)$ is then constructed by averaging the correlation functions over all $N$ particles of the system. 

\end{document}